\documentclass[11pt]{article}

\usepackage[utf8]{inputenc}
\usepackage[T1]{fontenc}
\usepackage{amsmath}
\usepackage{amssymb}
\usepackage{amsfonts}
\usepackage{amsthm}
\usepackage{mathtools}
\usepackage{mathrsfs}
\usepackage{bm}
\usepackage{booktabs}
\usepackage{array}
\usepackage{enumitem}
\usepackage{graphicx}
\usepackage[dvipsnames]{xcolor}
\usepackage[final,expansion=false]{microtype}
\usepackage[a4paper,top=2.5cm,bottom=2.5cm,left=3cm,right=3cm]{geometry}
\usepackage[colorlinks=true,linkcolor=MidnightBlue,citecolor=MidnightBlue,
            urlcolor=MidnightBlue]{hyperref}

\allowdisplaybreaks
\numberwithin{equation}{section}
\setlist[enumerate]{leftmargin=*,itemsep=0.4em}
\setlist[itemize]{leftmargin=*,itemsep=0.25em}

\theoremstyle{plain}
\newtheorem{theorem}{Theorem}[section]

\newtheorem{lemma}[theorem]{Lemma}

\theoremstyle{definition}

\theoremstyle{remark}
\newtheorem{remark}[theorem]{Remark}

\newcommand{\dd}{\mathop{}\!\mathrm d}

\newcommand{\Lie}{\mathcal L}
\newcommand{\Rc}{\operatorname{Ric}}

\newcommand{\tr}{\operatorname{tr}}
\newcommand{\tf}{\operatorname{tf}}

\newcommand{\Boxg}{\Box_{\mathbf g}}
\newcommand{\norm}[1]{\left\lVert #1\right\rVert}
\newcommand{\abs}[1]{\left\lvert #1\right\rvert}

\newcommand{\bg}{\mathbf g}
\newcommand{\slg}{\gamma}
\newcommand{\roundg}{\overset{\circ}{\gamma}}

\newcommand{\snab}{\nabla\mkern-13mu /\,\,}
\newcommand{\sdiv}{\mbox{div}\mkern-19mu /\,\,\,\,}

\newcommand{\Hb}{\underline H}

\newcommand{\chib}{\underline\chi}
\newcommand{\chih}{\widehat\chi}
\newcommand{\chibh}{\widehat{\underline\chi}}
\newcommand{\etab}{\underline\eta}
\newcommand{\omegab}{\underline\omega}

\newcommand{\betab}{\underline\beta}

\newcommand{\dv}{\operatorname{div}}
\newcommand{\cl}{\operatorname{curl}}

\def\oc{\overset{\circ}}
\def\wt{\widetilde}
\hypersetup{
  pdftitle={Picard Iteration for the Characteristic Initial Value Problem in Einstein Equations},
  pdfauthor={Shengrong Wu}
}

\title{Picard Iteration for the Characteristic Initial Value Problem in Einstein Equations}
\author{Shengrong Wu
\thanks{Department of Mathematics,
National University of Singapore.
Email: \href{mailto:shengrong_wu@u.nus.edu}
{\texttt{shengrong\_wu@u.nus.edu}}}}
\date{}

\begin{document}

\maketitle
\begin{abstract}
    We present an iteration algorithm for vacuum and
    Einstein scalar-field equations in double-null gauge, which transform the non-linear PDE into systems of ODE. The numerical
    realization combines characteristic constraint solves, LGL spectral
    elements, pole-free spherical operators, Galerkin projection, and
    independent first-order residual and consistency checks.
\end{abstract}
\tableofcontents

\section{Introduction}

The characteristic initial value problem is a natural formulation of the
Einstein equations when radiation, null focusing, and the formation of
trapped surfaces are central.  Instead of prescribing data on a spacelike
hypersurface, one prescribes compatible data on two intersecting null
hypersurfaces and evolves their future domain of dependence.  Local
well-posedness for this problem is known under suitable regularity
assumptions~\cite{rendall-90,luk-12,cabet-16,hilditch-20}, but its direct
numerical realization becomes substantially more difficult once spherical
symmetry is removed.  The use of null hypersurfaces as an evolution framework
goes back to the Bondi--Sachs description of gravitational radiation and the
early analysis of characteristic data~\cite{bondi-62,sachs-civp-62}; its
development as a numerical method is surveyed in~\cite{winicour-12}.

Characteristic numerical relativity has progressed from double-null vacuum
evolutions with two Killing symmetries~\cite{corkill-stewart-83} and
spherically symmetric scalar collapse in null coordinates
~\cite{garfinkle-95,hamade-stewart-96} to axisymmetric vacuum evolution on
outgoing null cones~\cite{gomez-94} and generic angular data for accurate
Cauchy--characteristic waveform extraction~\cite{moxon-23}.  Numerical
relativity has also long evolved nonspherical spacetimes in spacelike
foliations.  Fully three-dimensional Cauchy calculations have studied
scalar-field collapse without symmetry assumptions~\cite{deppe-19}, and
axisymmetric pseudospectral calculations have reached strongly aspherical
regimes in which the center of collapse bifurcates~\cite{marouda-24}.  The
first null-coordinate simulations of nonspherical regular data collapsing
to a black hole were obtained in twist-free axisymmetry for the massless
scalar field~\cite{gundlach-24}; that work describes the accessible data as
moderately nonspherical and identifies larger deviations from spherical
symmetry, as well as vacuum collapse, as important challenges.  Thus the gap
addressed here is not an absence of nonspherical numerical solutions in
general.  It is the relative scarcity of direct characteristic evolutions
with strongly nonspherical, freely prescribed initial geometry in a fully
angular double-null gauge.

Let $(u,v,\theta^A)$ be double-null coordinates.  We write the spacetime
metric as
\begin{equation}\label{eq:introduction-double-null-metric}
  \bg=-2\Omega^2(\dd u\otimes\dd v+\dd v\otimes\dd u)
  +g_{AB}(\dd\theta^A-b^A\dd u)\otimes
   (\dd\theta^B-b^B\dd u).
\end{equation}
The metric is therefore represented by the section metric $g_{AB}$, the
lapse $\Omega$, and the angular shift $b$.  This formulation gives
considerable freedom to choose the geometry on the two initial null
hypersurfaces, subject to the characteristic constraints and corner
compatibility.  In particular, the nonspherical part of the initial metric
variation can be prescribed directly rather than produced only indirectly
from a matter perturbation or a spacelike constraint solve.  The fourth
experiment gives a quantitatively strong example: on its outgoing initial
hypersurface, where $\Omega=1$, the prescribed data satisfy
\begin{equation}\label{eq:introduction-strong-asymmetry}
  \norm{\tf\Lie_{\partial_v}g(-1,0.005)}_{L^2(S)}=7.50960,
  \qquad
  \norm{\tf\Lie_{\partial_v}g(-1,0.005)}_{L^\infty(S)}=2.53468.
\end{equation}
The perturbation is consequently a substantial geometric departure from
spherical symmetry, rather than a nonspherical profile whose effect on the
initial metric is numerically negligible.

This paper develops a first-order Picard iteration for both the Einstein
vacuum equations
\begin{equation}
  \Rc(\bg)=0
\end{equation}
and the Einstein--massless-scalar system
\begin{equation}
  \Rc(\bg)=\dd\phi\otimes\dd\phi,
  \qquad \Boxg\phi=0,
\end{equation}
in the gauge \eqref{eq:introduction-double-null-metric}.  The iteration is
organized by the geometry of the characteristic problem.  Data that are
free, data fixed by the null constraints, and quantities transported in the
two characteristic directions play different roles in each sweep.  The
resulting construction is therefore not obtained by treating all coordinate
components as an undifferentiated system of equations: its update order
follows the causal and geometric dependency structure of the double-null
equations.

The iteration architecture is adapted from the approximate-spacetime
construction introduced by An and Wu~\cite{wu-26}.  Both constructions use a
triangular hierarchy of double-null transport equations, with nonlinear
coefficients evaluated at the preceding iterate.  The present work modifies
that analytic construction into an executable numerical Picard map: it uses
numerically convenient weighted variables and equivalent transport
identities, imposes both characteristic traces discretely, and incorporates
projection, relaxation, and residual auditing.  Thus the underlying
geometric organization is largely the same, while some update equations and
terms are reformulated for numerical computation.

Two difficulties are particularly important.  The first is to obtain an
executable form of the equations.  Analytic arguments can often group
lower-order terms schematically or pass between conventionally equivalent
forms, whereas numerical evolution requires every displayed sign and
numerical factor to be fixed consistently.  We address this by checking the
formulas against identities derived from the four-metric, by testing exact
solutions, and by evaluating first-order residuals using fresh derivatives
of the stored weighted fields rather than the construction sources of a
Picard sweep.  We check the metric--connection consistency separately and
avoid second null derivatives at the rough characteristic endpoint.  The second difficulty is the
iteration itself.  Its design requires identifying a closed hierarchy that
respects the characteristic constraints, transports information in the
correct direction, preserves the two initial traces, and remains meaningful
as the null hypersurfaces focus.  This construction comes from the geometric
structure of the equations rather than from a generic fixed-point template.

The numerical implementation combines Legendre--Gauss--Lobatto spectral
elements in the two null coordinates with pole-free spherical
differentiation, angular Galerkin projection, characteristic constraint
solves, and independent overgrid audits.  Extensive AI-assisted software
development made it practical to implement, refactor, and test this large
coupled system.  The mathematical conventions, iteration design, experiment
definitions, acceptance criteria, and scientific interpretation remain
author-controlled; confidence in the computations is based on exact
benchmarks, convergence studies, immutable characteristic data, and
independent residual evaluation rather than on code generation itself.

Eight numerical experiments test complementary aspects of the method.  The
exact-solution tests cover Schwarzschild, Kerr, and Fisher--JNW spacetimes.
They measure the metric error in regular regions, across a Schwarzschild
horizon, and toward a curvature singularity.  The vacuum experiments evolve a
strong nonspherical outgoing perturbation and crossed characteristic data
whose initial curvature is singular at the corner.  The final experiments
evolve nonspherical Einstein--scalar data and a stronger scalar pulse toward
a trapped region.  In the last experiment, an inner atlas of characteristic
rectangles covers a curved domain adapted to null focusing.  Independent
four-metric evaluation finds the section
\begin{equation}\label{eq:introduction-trapped-section}
  (u,v)=(-0.4696271025,0.04)
\end{equation}
on which the two future null expansions are strictly negative everywhere,
with respective spherical suprema $-0.0802152$ and $-4.28011$.  Solving the
angular marginally outer trapped surface equation on successive incoming
null cones then reconstructs an apparent-horizon tube of 18 sections over
$0.00441\leq v\leq0.05$.  Coordinate and angular controls agree on its
$v=0.04$ section to better than $1.6\times10^{-7}$ in the graph location.

The trapped-region and MOTS computations are motivated by An's perturbation
construction and by Roesch and Scheuer's mean curvature flow in null
hypersurfaces.  An proves that anisotropic outgoing characteristic
perturbations of Christodoulou's naked-singularity data generate an
anisotropic apparent horizon that censors the singularity~\cite{an-25}.
Roesch and Scheuer show, under suitable geometric assumptions, that mean
curvature flow within a null hypersurface converges smoothly from an outer
untrapped initial surface to a MOTS~\cite{roesch-scheuer-22}.  Here the flow
result provides geometric motivation for MOTS detection; the numerical
implementation solves the angular MOTS graph equation directly by nonlinear
least squares.  The experiment does not constitute a proof of either
theorem, and the reconstructed tube is tied to the chosen incoming-null
foliation.  It provides a coordinate- and angularly controlled numerical
realization of the trapped-region mechanism and its anisotropic
apparent horizon.

Section~2 fixes the double-null equations and records the analytic framework.
Section~3 constructs the Picard map and its numerical discretization.
Sections~4 and~5 present the vacuum and Einstein--scalar experiments,
respectively, and distinguish exact metric errors from independently
evaluated curvature residuals.
\section{Equations and Theoretical Analysis}\label{sec:theory}
The null-frame decomposition of the Einstein equations and the associated
geometric energy estimates were developed systematically in the proof of the
nonlinear stability of Minkowski space~\cite{christodoulou-klainerman-93}.
Klainerman and Nicol\`o subsequently formulated the exterior evolution problem
using a double-null foliation~\cite{klainerman-nicolo-03}, while
Christodoulou's treatment of black-hole formation develops the optical
structure equations, characteristic initial data, and null estimates in the
same geometric setting~\cite{christodoulou-09}.  This section fixes the
particular conventions used in the numerical construction and then recalls
the local existence framework relevant to the characteristic problem.
\subsection{Double null foliation and equations}
In this section, we introduce the equations for Lorentzian metric under double null foliation. Let $(\mathcal{M},g)$ be a $4$-dimensional Lorentzian manifold. With double null coordinates $(u,v,\theta^A)$, we use $(g_{AB},b^A,\Omega)$ to represent the metric \eqref{eq:introduction-double-null-metric}. The level sets of $u$ and $v$ are denoted by $H_u$ and $\Hb_{v}$ respectively, and their spherical intersection $H_u\cap\Hb_v$ is denoted by $S_{u,v}$.
With double null frame 
\[e_3=\Omega^{-1}(\partial_u+b),\ e_4=\Omega^{-1}\partial_v,\ e_A=\partial_{\theta^A},\]
we can define Ricci coefficients, or connection components
\begin{equation}\label{Ricci_coefficients_definition}
    \begin{aligned}
    \chi_{A B}=g\left(D_A e_4, e_B\right), &\quad \underline{\chi}_{A B}=g\left(D_A e_3, e_B\right), \quad
    \eta_A=-\frac{1}{2} g\left(D_3 e_A, e_4\right),\\
    \underline{\eta}_A=-\frac{1}{2} g\left(D_4 e_A, e_3\right),&\quad 
    {\omega}=-\frac{1}{4} g\left(D_4 e_3, e_4\right), \quad \underline{\omega}=-\frac{1}{4} g\left(D_3 e_4, e_3\right),\\
    \zeta_A=\frac{1}{2}g\left(D_A e_4,e_3\right)&=-\frac{1}{4}\Omega^{-2}(\Lie_vb^B)g_{AB}.
    \end{aligned}
\end{equation}
Their transport equations are listed below:
\begin{equation}\label{eq:Ricci_coefficients_propagation_1}
    \begin{aligned}
        \nabla_4 \operatorname{tr} \chi+|{\chi}|^2 =&-\operatorname{Ric}_{44}-2 \omega \operatorname{tr} \chi, \quad 
     \nabla_3 \operatorname{tr} \underline{\chi}+|\underline{{\chi}}|^2 =-\operatorname{Ric}_{33}-2 \underline{\omega} \operatorname{tr} \underline{\chi},
    \end{aligned}
\end{equation}
\begin{equation}\label{eq:Ricci_coefficients_propagation_2}
    \begin{aligned}
        \nabla_3 \hat{\chi}_{A B}+\frac{1}{2} \operatorname{tr} \underline{\chi} \hat{\chi}_{A B}=&\widehat{\operatorname{Ric}}_{A B}+2 \underline{\omega} \hat{\chi}_{A B}+(\snab  \hat{\otimes} \eta)_{A B}+(\eta \hat{\otimes} \eta)_{A B}-\frac{1}{2} \operatorname{tr} \chi \underline{\hat{\chi}}_{A B}, \\
     \nabla_4 \underline{\hat{\chi}}_{A B}+\frac{1}{2} \operatorname{tr} \chi \underline{\hat{\chi}}_{A B}=&\widehat{\operatorname{Ric}}_{A B}+2 \omega \underline{\hat{\chi}}_{A B}+(\snab\hat{\otimes} \underline{\eta})_{A B}+(\underline{\eta} {\hat{\otimes}} \underline{\eta})_{A B}-\frac{1}{2} \operatorname{tr} \underline{\chi} \hat{\chi}_{A B},
    \end{aligned}
\end{equation}
\begin{equation}
    \begin{aligned}
        \nabla_3 \operatorname{tr} \chi 
        =&-\tr\chi\tr\chib+2\omegab\tr\chi+2\sdiv\eta+2\left|\eta\right|^2-2K+R+\Rc_{34}, \\
        \nabla_4 \operatorname{tr} \underline{\chi} 
        =&-\tr\chi\tr\chib+2\omega\tr\chib+2\sdiv\etab+2\left|\etab\right|^2-2K+R+\Rc_{34}.
    \end{aligned}
\end{equation} 
\begin{equation}\label{eq:Ricci_coefficients_propagation_3}
    \begin{aligned}
        \nabla_4 \eta=&-\chi \cdot(\eta-\underline{\eta})-\frac{1}{2} R_{A 434}, \quad
        \nabla_3 \underline{\eta}=-\underline{\chi} \cdot(\underline{\eta}-\eta)-\frac{1}{2} R_{A 343}, \\
        \nabla_4 \underline{\omega} 
        =&2\omega\omegab-\frac{1}{2}  K+\frac{1}{2} \Rc_{34}+\frac{1}{4} R +\frac{1}{4}\chih\cdot\chibh-\frac{1}{8}\tr\chi\,\tr\chib+\frac{1}{2}\left|\eta\right|^2-\eta\cdot\etab, \\
        \nabla_3 \omega 
        =&2\omega\omegab-\frac{1}{2}  K+\frac{1}{2} \Rc_{34}+\frac{1}{4} R +\frac{1}{4}\chih\cdot\chibh-\frac{1}{8}\tr\chi\,\tr\chib+\frac{1}{2}\left|\underline\eta\right|^2-\eta\cdot\etab.
    \end{aligned}
\end{equation}
Moreover, we have equations for $\zeta$:
\begin{equation}\label{eq:Ricci_coefficients_propagation_4}
    \begin{aligned}
        &\Omega\Rc_{3A}=\Omega\nabla_3\zeta+\frac{3}{2}\Omega\tr\chib\zeta+\Omega\chibh\cdot\zeta+\Omega\tr\chib\nabla\log\Omega\\
        &\qquad\qquad\qquad+2\nabla(\Omega\omegab)+\dv(\Omega\chibh)-\frac{1}{2}\nabla(\Omega\tr\chib),\\
        &\Omega\Rc_{4A}=-\Omega\nabla_4\zeta-\frac{3}{2}\Omega\tr\chi\zeta-\Omega\chih\cdot\zeta+\Omega\tr\chi\nabla\log\Omega\\
        &\qquad\qquad\qquad+2\nabla(\Omega\omega)+\dv(\Omega\chih)-\frac{1}{2}\nabla(\Omega\tr\chi).
    \end{aligned}
\end{equation}
% Here $K$ stands for Gaussian curvature of topological sphere $S_{u,v}$. Gauss-Codazzi equations imply 
% \begin{equation}\label{eq:Gauss-Codazzi_1}
%         \begin{aligned}
%          K=&\frac{1}{2} \sg^{A B} R_{3 A 4 B}+\frac{1}{2} R+\frac{1}{2} \operatorname{Ric}_{34}+\frac{1}{2} \hat{\chi} \cdot \underline{\hat{\chi}}-\frac{1}{4} \operatorname{tr} \chi \operatorname{tr} \underline{\chi}, 
%         \end{aligned}
%         \end{equation}
% \begin{equation}\label{eq:Gauss-Codazzi_2}
%             \begin{aligned} 
%             \snab^B \hat{\chi}_{A B}-\frac{1}{2} \snab_A \operatorname{tr} \chi=&-\frac{1}{2} R_{A 434}+\operatorname{Ric}_{4 A}+\frac{1}{2} \operatorname{tr} \chi \zeta_A-\zeta^B \hat{\chi}_{A B}, \\
%             \snab^B \underline{\hat{\chi}}_{A B}-\frac{1}{2} \snab_A \operatorname{tr} \underline{\chi}=&-\frac{1}{2} R_{A 343}+\operatorname{Ric}_{3 A}-\frac{1}{2} \operatorname{tr} \underline{\chi} \zeta_A+\zeta^B \underline{\hat{\chi}}_{A B}.
%             \end{aligned}
%         \end{equation}
% \begin{equation}\label{eq:curl-eta-etab}
%             \begin{aligned}
%                 \snab_A \eta_B-\snab_B \eta_A=&R_{4[A B] 3}+\underline{\hat{\chi}}^C{ }_{[A} \hat{\chi}_{B] C}, \\
%          \snab_A \underline{\eta}_B-\snab_B \underline{\eta}_A=&-R_{4[A B] 3}-\underline{\hat{\chi}}^C{ }_{[A} \hat{\chi}_{B] C}, \\
%             \end{aligned}
%         \end{equation}
For Weyl curvature,
\begin{equation*}
    W_{\mu\nu\theta\lambda}=R_{\mu\nu\theta\lambda}-\frac{1}{2}\left(\Rc_{\mu\theta}g_{\nu\lambda}+g_{\mu\theta}\Rc_{\nu\lambda}-\Rc_{\nu\theta}g_{\mu\lambda}-g_{\nu\theta}\Rc_{\mu\lambda}\right)+\frac{R}{6}\left(g_{\mu\theta}g_{\nu\lambda}-g_{\nu\theta}g_{\mu\lambda}\right),
\end{equation*}
we define the curvature components
\begin{equation}
\begin{aligned}
\beta_A & =\frac{1}{2} W\left(e_A, e_4, e_3, e_4\right), & & \underline{\beta}_A=\frac{1}{2} W\left(e_A, e_3, e_3, e_4\right), \\
\rho & =\frac{1}{4} W\left(e_4, e_3, e_4, e_3\right), & & \sigma=\frac{1}{4}{ }^* W\left(e_4, e_3, e_4, e_3\right).
\end{aligned}
\end{equation}
Then Gauss-Codazzi equations imply the following algebraic relations,
\begin{equation}
    K=-\rho+\frac{1}{3}R+\frac{1}{2}\Rc_{34}+\frac{1}{2}\chibh\cdot\chih-\frac{1}{4}\tr\chib\tr\chi,
\end{equation}
\begin{equation}
    \cl\eta=-\cl\etab=\sigma+\frac{1}{2}\chibh\wedge\chih,
\end{equation}
\begin{equation}\label{Codazzi_eq_chi}
    \dv\chih-\frac{1}{2}\nabla\tr\chi=-\beta+\frac{1}{2}\Rc_{4\cdot}+\frac{1}{2}\tr\chi\zeta-\zeta\cdot\chih,
\end{equation}
\begin{equation}\label{Codazzi_eq_chib}
    \dv\chibh-\frac{1}{2}\nabla\tr\chib=\betab+\frac{1}{2}\Rc_{3\cdot}-\frac{1}{2}\tr\chib\zeta+\zeta\cdot\chibh.
\end{equation}
After introducing the reduced curvature
\begin{equation}
    \beta^r_A=\beta_A-\frac{1}{2}\Rc_{4A},\quad \betab^r_A=\betab_A+\frac{1}{2}\Rc_{3A}, \quad \sigma^r=\sigma+\frac{1}{2}\chibh\wedge\chih,
\end{equation}
we can write the equations for curvature components
\begin{equation}
    \begin{aligned}
        &\Omega\nabla_3 K+\Omega\tr\chib K=\dv \left(\Omega\betab^r\right)+\frac{1}{2}\dv\left(\eta\Omega\tr\chib\right)-\dv\left(\eta\cdot\Omega\chibh\right),\\
         &\Omega\nabla_4 K+\Omega\tr\chi K=-\dv \left(\Omega\beta^r\right)+\frac{1}{2}\dv\left(\etab\Omega\tr\chi\right)-\dv\left(\etab\cdot\Omega\chih\right),
    \end{aligned}
\end{equation}
\begin{equation}
    \begin{aligned}
        \nabla_3\beta^r&+\left(\tr\chib-2\omegab\right) \beta^r=-\nabla K+{}^*\nabla\sigma+2\chih\cdot\betab\\
        &\qquad+\frac{1}{2}\left(\nabla(\chih\cdot\chibh)-{}^*\nabla(\chih\wedge\chibh)\right)-\frac{1}{4}\nabla\left(\tr\chi\tr\chib\right)\\
        &\qquad+3(\eta\rho+{}^*\eta\sigma)+\nabla_A\left(g^{CD}\Rc_{CD}\right)-\nabla^B \Rc_{BA}-\frac{1}{2}\tr\chib \Rc_{4A},\\
        \nabla_4\betab^r&+\left(\tr\chi -2\omega\right)\betab^r=\nabla K+{}^*\nabla\sigma+2\chibh\cdot\beta\\
        &\qquad-\frac{1}{2}\left(\nabla(\chih\cdot\chibh)+{}^*\nabla(\chih\wedge\chibh)\right)+\frac{1}{4}\nabla\left(\tr\chi\tr\chib\right)\\
        &\qquad-3\left(\etab\rho-{}^*\etab\sigma\right) -\nabla_A(g^{CD}\Rc_{CD})+\nabla^B \Rc_{BA}+\frac{1}{2}\tr\chi \Rc_{3A}.
    \end{aligned}
\end{equation}
Using \eqref{eq:Ricci_coefficients_propagation_3}, we derive 
\begin{equation}
  \begin{aligned}
      \Omega\nabla_3\cl\etab+\Omega\tr\chib\cl\etab=&\cl(\Omega\betab^r)+\epsilon^{AC}\nabla_A\left(\Omega\chib\cdot\eta\right)_C-\epsilon^{AC}\nabla_A(\Omega\Rc)_{C3},\\
      \Omega\nabla_4\cl\eta+\Omega\tr\chi\cl\eta=&-\cl(\Omega\beta^r)+\epsilon^{AC}\nabla_A\left(\Omega\chi\cdot\etab\right)_C-\epsilon^{AC}\nabla_A(\Omega\Rc)_{C4},
  \end{aligned}
\end{equation}
and
\begin{equation}
    \begin{aligned}
        &\Omega\nabla_3\left(\dv \etab-K\right)+\Omega\tr\chib\left(\dv \etab-K\right)=4\dv(\Omega\chibh\cdot\zeta)-\dv(\Omega\Rc_3),\\
        &\Omega\nabla_4\left(\dv \eta-K\right)+\Omega\tr\chi\left(\dv \eta-K\right)=-4\dv(\Omega\chih\cdot\zeta)-\dv(\Omega\Rc_4).
    \end{aligned}
\end{equation}
For a scalar function $\phi$, there is identity
\begin{equation}\label{eq:ese-wave-eq}
    -\Omega^2\square\phi=\Omega e_3(\Omega e_4\phi)+\frac{1}{2}\Omega\tr\chi\Omega e_3\phi+\frac{1}{2}\Omega\tr\chib\Omega e_4\phi-\Omega^2\Delta\phi-2\Omega^2\eta\cdot\nabla\phi.
\end{equation}

\subsection{Local existence}
We now introduce some results on Einstein vacuum equations. Similar results can be derived for other Einstein field equations and we omit them due to length of the article.

For data prescribed on two transversely intersecting null hypersurfaces,
Rendall reduced the characteristic problem to a standard Cauchy problem and
proved existence in a neighborhood of the intersection
~\cite{rendall-90}.  Luk then extended the solution to a neighborhood of the
full initial null hypersurfaces by estimates adapted to a double-null
foliation~\cite{luk-12}.  Related existence results have since been obtained
for nonlinear symmetric hyperbolic systems, including Einstein equations with
sources~\cite{cabet-16}, and in the Newman--Penrose formalism in Stewart's
gauge~\cite{hilditch-20}.  We state the vacuum theorem in the form used here,
following~\cite{luk-12}.
\begin{theorem}
    Given initial null hypersurface $H_0,\Hb_0$ and regular initial metric $(g,\Omega,b)$ on it, we define $\chi_{AB}=\frac{1}{2}\Lie_{e_4}g_{AB}$ on $H_0$ and $\chib_{AB}=\frac{1}{2}\Lie_{e_3}g_{AB}$ on $\Hb_0$. Let $\zeta$ be a vector field on $H_{0}\cap \Hb_0$. Suppose 
    \[e_4(\tr\chi)+|\chi|^2+2\omega\tr\chi=0,\ e_3(\tr\chib)+|\chib|^2+2\omegab\tr\chib=0,\]
    hold on $H_0,\Hb_0$ respectively, then there is a future open neighborhood of $H_0\cup \Hb_0$ and metric $(g,b,\Omega)$ on it solving the Einstein vacuum equations and extending the initial data $(g,\Omega,b)$ and $\zeta$. 
\end{theorem}
Thus local existence extends along every portion of the initial null
hypersurfaces on which the required norms remain controlled.  Extension to a
larger rectangle, such as $(0,1)^2\times\mathbb{S}^2$, requires corresponding
a priori bounds throughout that region.

\subsection{Residual control}
In this paper, we target to deal with asymmetric Einstein equations, which in general do not have explicit solutions. To measure the computational error, we use Ricci residues. For Einstein equation 
\[G_{\mu\nu}:=\Rc_{\mu\nu}-\frac{1}{2}Rg_{\mu\nu}=T_{\mu\nu},\]
we measure $\|G-T\|_{L^2(S_{u,v})}$ to evaluate the approximation. The theorems below reveals that in our case the difference of metrics is controlled by the difference of curvature residues.

\begin{theorem}
    We consider metric $\oc{g}_{AB},\oc{b},\oc{\Omega}$ on $(0, 1)\times (0,1)\times \mathbb{S}^2$ such that
    \[\oc{\Rc}_{44}(0,v)=0,\ \oc{\Rc}_{33}(u,0)=0,\ \oc{\Rc}_{4A}(0,v)=0,\ \oc{\Rc}_{3A}(u,0)=0,\]
    % \[\oc{\Rc}_{AB}(u,0)=0,\ \oc{\Rc_{34}}(u,0)=0,\ \oc{R}(u,0)=0,\]
    and estimates 
    \begin{equation}
        \|\oc{\Rc}_{AB},\oc{\Rc}_3,\oc{\Rc}_4,\oc{\Rc}_{34},\oc{\Rc}_{33},\oc{\Rc}_{44}\|_{L^\infty_uL^\infty_vH^3(S)}\leq \epsilon.
    \end{equation}
    We assume moreover 
    \begin{equation}
        \| \partial_{\mathbb{S}}\oc{g}, \partial_{\mathbb{S}}\oc{b}, \partial_{\mathbb{S}}\oc\Omega, \oc{\psi}\|_{L^\infty_u L^\infty_v H^4(S)} \leq C_0,
    \end{equation}
    where $\psi$ stands for Ricci coefficients in interest.
    Then for $\epsilon=\epsilon(C_0)$ sufficiently small, there exists $(g,b,\Omega)$ solving EVE on $( 0,1)\times (0,1)\times \mathbb{S}^2$ with initial data 
    \[(g,b,\Omega)=(\oc{g},\oc{b},\oc\Omega),\ {\rm on}\ \Hb_0\cup H_{0},\]
    and estimates
    \begin{equation}
        \|g-\oc{g},b-\oc{b},\Omega-\oc{\Omega}\|_{L^\infty_u L^\infty_v H^4(S)}\leq \epsilon^{1/2}.
    \end{equation}
\end{theorem}

The methods of deriving the equations of difference and doing estimates are similar to section 7 of \cite{an-26},
so we only write the proof in sketch. 

Define auxiliary functions $\omega^\dagger,\omegab^\dagger$ by
\[\nabla_3\omega^\dagger=\frac{1}{2}\cl\zeta,\quad \omega^\dagger|_{H_0}=0,\quad \nabla_4\omegab^\dagger=\frac{1}{2}\cl\zeta,\quad \omegab^\dagger|_{\Hb_0}=0.\]
We make bootstrap assumption
\begin{equation*}
    \begin{aligned}
        &\|\wt{\beta},\wt{K},\nabla\wt\eta,\nabla\wt{\etab},\nabla\wt{\chi},\nabla\wt{\omega},\nabla\wt{\omega^\dagger}\|_{L^2_vH^3(S)}\\
        &\qquad+\|\wt{K},\wt{\betab},\nabla\wt{\eta},\nabla\wt\etab,\nabla\wt{\chib},\nabla\wt{\omegab},\nabla\wt{\omegab^\dagger}\|_{L^2_uH^3(S)}\leq B\exp(D(v+u))\epsilon.
    \end{aligned}
\end{equation*}
We first fix $B$ sufficiently large to control the initial-data and forcing terms, and then choose $D$ sufficiently large to improve the bootstrap bound from $B$ to $B/2$.

For motivation, let $f\geq0$ satisfy $\partial_uf\leq C_0f$ and $\partial_vf\leq C_0f$. Assume its data on $u=0$ and $v=0$ are bounded by $C_0\epsilon$. Under the bootstrap bound
$$f(u,v)\leq B\exp(D(v+u))\epsilon,$$ integration gives
\begin{equation*}
    \begin{aligned}
        f(u,v)-f(0,v)\leq&C_0\int_{0}^u f(u^\prime,v) du^\prime\\
        \leq  & \frac{C_0B}{D}\left(\exp(D(u+v))-\exp(Dv)\right)\epsilon \leq \frac{C_0B}{D} \exp(D(u+v))\epsilon,\\
        f(u,v)-f(u,0)\leq & C_0\int_{0}^v f(u,v^\prime) dv^\prime\\
        \leq & \frac{C_0B}{D}\left(\exp(D(u+v))-\exp(Du)\right)\epsilon \leq \frac{C_0B}{D} \exp(D(u+v))\epsilon.
    \end{aligned}
\end{equation*}
Choosing $B\geq4C_0$ and $D\geq4C_0$ improves this bound to $(B/2)\exp(D(u+v))\epsilon$. With $B$ and $D$ fixed independently of $\epsilon$, the resulting bound is $O(\epsilon)=o(\epsilon^{1/2})$ on a fixed finite region.

For the energy estimates we use the following integration bounds. All null integrals run from $0$ to the indicated endpoint, and the hypotheses hold throughout the rectangle.
\begin{lemma}
    Let $D>0$. Suppose $\|f\|_{L^2(S_{u,v})}\leq B\exp(D(v+u))\epsilon$, then we have
    \begin{equation}
        \|f\|_{L^2_uL^2(S)}+\|f\|_{L^2_vL^2(S)} \leq \sqrt{2}BD^{-1/2}\exp(D(v+u))\epsilon.
    \end{equation}

    Suppose either $\|h\|_{L^2_uL^2(S)}\leq B\exp(D(u+v))\epsilon$ or $\|h\|_{L^2_vL^2(S)}\leq B\exp(D(u+v))\epsilon$, then we have    
    \begin{equation}
        \|h\|_{L^2_uL^2_vL^2(S)} \leq B(2D)^{-1/2}\exp(D(v+u))\epsilon.
    \end{equation}
    The same estimates hold with $H^3(S)$ in place of $L^2(S)$.
\end{lemma}

We use notation
\[\psi_1\in\{\eta,\etab,\chi,\omega\},\ \psi_2\in\{\eta,\etab,\chib,\omegab\}\]
Because 
\[\Lie_v b=-2\Omega^2(\eta-\etab),\ \Lie_vg=2\Omega\chi,\ \Lie_v\log\Omega=-2\Omega\omega,\]
we obtain that
\begin{equation}
    \|\wt{b},\wt{g},\wt{\log\Omega}\|_{H^4(S_{u,v})}\lesssim \|\wt{\eta},\wt{\etab},\wt{\chi},\wt{\omega},\wt{\Omega},\wt{g}\|_{L^2_vH^4(S)}\lesssim B\exp(D(u+v))\epsilon.
\end{equation}
For Bianchi pair $(\Psi_1,\Psi_2)$ in interest, we consider equations 
\begin{equation*}
    \begin{aligned}
        \Omega\nabla_3\wt{\Psi_1} &+\mathcal{D}\wt{\Psi_2}=\wt{\psi\Psi}+\wt{\psi\nabla\psi}+\wt{\psi^3}+\wt{b}\cdot\nabla\oc{\Psi_1}+\wt{\partial g}\oc{\Psi_2}+\wt{g}\oc{\nabla\Psi_2}+\oc\psi\oc\Rc+\oc{\nabla\Rc},\\
        \Omega\nabla_4\wt{\Psi_2} &-{}^*\mathcal{D}\wt{\Psi_1}=\wt{\psi\Psi}+\wt{\psi\nabla\psi}+\wt{\psi^3}+\wt{\partial g}\oc{\Psi_1}+\wt{g}\oc{\nabla\Psi_1}+\oc\psi\oc\Rc+\oc{\nabla\Rc}.
    \end{aligned}
\end{equation*}
Energy estimates give that 
\begin{equation*}
    \begin{aligned}
        &\|\wt\Psi_1\|^2_{L^2_vH^3(S)}+\|\wt\Psi_2\|^2_{L^2_uH^3(S)}\\
        &\qquad\qquad\lesssim \epsilon^2+\|\wt{\partial g},\wt{g},\wt{\psi},\wt{\nabla\psi},\wt\Psi,\wt{b}\|^2_{L^2_uL^2_vH^3(S)}\lesssim (1+B^2/D)\exp(2D(u+v))\epsilon^2.
    \end{aligned}
\end{equation*}
Applying the estimate to pairs
\[(\beta,(K,\cl\eta)),\ ((K,\cl\etab),\betab),\ (\nabla\tr\chib,0),\ (0,\nabla\tr\chi),\]
\[ (\dv\etab-K,0),\ (0,\dv\eta-K),\ (\nabla\omega+{}^*\nabla\omega^\dagger-\frac{1}{2}\beta,0),\ (0,-\nabla\omegab+{}^*\nabla\omegab^\dagger-\frac{1}{2}\betab),\]
yields a bound $C(1+B^2/D)\exp(2D(u+v))\epsilon^2$ for the square of the full bootstrap norm, where $C=C(C_0)$ is independent of $B$, $D$, and $\epsilon$. Choosing $B^2\geq8C$ and then $D\geq8C$ improves the bootstrap bound to $(B/2)\exp(D(u+v))\epsilon$, completing the proof sketch.
\begin{remark}
    The conditions 
    \[\oc{\Rc}_{44}(0,v)=0,\ \oc{\Rc}_{33}(u,0)=0,\]
    guarantee that $(\oc{g},\oc{b},\oc{\Omega})$ is a valid initial data set for characteristic initial value problem of Einstein vacuum equation, and 
    \[\oc{\Rc}_{4A}(0,v)=0,\ \oc{\Rc}_{3A}(u,0)=0,\]
    ensure that $\oc\zeta$ along $H_{0}$ and $\Hb_0$ is equal to the $\zeta$ of corresponding genuine solution. 
\end{remark}

\section{Picard Iteration Design and Numerical Strategies}
In this section, we introduce the iteration design and numerical strategies.
The Picard hierarchy below is a numerical adaptation of the approximation
scheme in~\cite{wu-26}.  We preserve its principal dependency order---freezing
nonlinear coefficients at the previous iterate and successively solving
transport equations in their geometrically distinguished null directions---but
modify individual equations when an equivalent form is more suitable for
numerical integration.  Accordingly, the formulas in this section define the
numerical iteration used in the present work; they should not be read as a
literal transcription of the analytic scheme in~\cite{wu-26}.

\subsection{Picard iteration for Einstein equations}
In this section, we first consider Einstein scalar-field equations
\[\Rc=d\phi\otimes d\phi,\ \square_g\phi =0.\]

We work in region 
$$Q=\{-1<u<u_0,\ 0<v<v_0(u)\}.$$ 
The first task is to construct a valid initial data for ESE.
We directly prescribe $g,b,\Omega$ along $\Hb_0$. The only requirement is 
\[-\Omega\nabla_3(\Omega\tr\chib)-\frac{1}{2}(\Omega\tr\chib)^2-\left|\Omega\chibh\right|^2-4\Omega\omegab\Omega\tr\chib\geq 0\]
Then $\Omega e_3\phi(u,0)$ can be assigned via equation
\[\Omega e_3\phi=\left(-\Omega\nabla_3(\Omega\tr\chib)-\frac{1}{2}(\Omega\tr\chib)^2-\left|\Omega\chibh\right|^2-4\Omega\omegab\Omega\tr\chib\right)^{1/2}.\]
After defining $\phi(-1,0)$ and $\zeta(-1,0)$, we derive $\phi(u,0)$ from $\Omega e_3\phi(u,0)$ and $\zeta(u,0)$ via
\[\Omega\nabla_3\zeta+\frac{3}{2}\Omega\tr\chib\zeta+\Omega\chibh\cdot\zeta=-2\nabla(\Omega\omegab)-\dv(\Omega\chibh)+\frac{1}{2}\nabla(\Omega\tr\chib)-\Omega\tr\chib\nabla\log\Omega+\Omega e_3\phi\nabla\phi.\]
Thus $\eta=\zeta+\nabla\log\Omega$ and $\etab=-\zeta+\nabla\log\Omega$ are also prescribed.
We then define $\Omega\tr\chi(-1,0)$ and $\Omega\chih(-1,0)$, and compute $\Omega\chi(u,0)$ by 
\[\Omega\nabla_3(\Omega\tr\chi)+\Omega\tr\chib\Omega\tr\chi=\Omega^2\left(2\dv\eta+2\left|\eta\right|^2-2K(g)+|\nabla\phi|^2\right),\]
\[\Omega\nabla_3(\Omega\chih)+\frac{1}{2}\Omega\tr\chib\,\Omega\chih=\Omega^2\left(\nabla\hat\otimes\eta+\eta\hat\otimes\eta\right)-\frac{1}{2}\Omega\tr\chi\,\Omega\chibh+\frac{1}{2}\Omega^2\nabla\phi\hat\otimes\nabla\phi.\]
We also directly assign $\Omega e_4\phi(-1,0)$ and $\Omega\omega(-1,0)$, and compute them along $\Hb_0$ via
\[\Omega e_3(\Omega e_4\phi)+\frac{1}{2}\Omega\tr\chi\Omega e_3\phi+\frac{1}{2}\Omega\tr\chib\Omega e_4\phi-\Omega^2\Delta\phi-2\Omega^2\eta\cdot\nabla\phi=0,\]
\[\frac{1}{2}\Omega e_3\phi\Omega e_4\phi+\frac{1}{2}\Omega^2\left|\nabla\phi\right|^2=2\Omega\nabla_3(\Omega\omega)+\Omega^2K-\frac{1}{2}\Omega\chih\cdot\Omega\chibh+\frac{1}{4}\Omega\tr\chi\Omega\tr\chib-\Omega^2|\etab|^2+2\Omega^2\eta\cdot\etab.\]
Along $H_{-1}$, we prescribe $\Omega e_4\phi(-1,v)$ and $\Omega\omega(-1,v)$ to compute $\phi,\Omega$. Consider a symmetric and $g(-1,0)$-trace-free tensor $X_{AB}(-1,v)$.
We then prescribe $g,\Omega\chi$ along $H_{-1}$ by solving:
\begin{equation}
    \left\{\begin{aligned}
         & \Omega\chih_{AB}=\frac{1}{2}\left(X_{AC}\left(g(-1,0)\right)^{CD}{g}_{DB}+X_{BC}\left(g(-1,0)\right)^{CD}{g}_{DA}\right),\\
        &\partial_v (\Omega\tr\chi)=-\frac{1}{2}{\left(\Omega\tr\chi\right)}^2-4\Omega\omega\Omega\tr\chi-(\Omega e_4\phi)^2-\Omega\chih_{AB}\Omega\chih_{CD}g^{AC}g^{BD},\\
        &\Lie_v g_{AB}=\Omega\tr\chi g_{AB}+2\Omega\chih_{AB}.
    \end{aligned}\right.
\end{equation}

With these initial data, we can start the iteration.
The quantities to be iterated are 
\[g^{(i)},(b^{(i)})^A,\log\Omega^{(i)},(\Omega\chi)^{(i)}_{AB},(\Omega\chib)^{(i)}_{AB},(\zeta^{(i)})^A,(\Omega\omega)^{(i)},(\Omega\omegab)^{(i)},(\Omega e_3\phi)^{(i)}, (\Omega e_4\phi)^{(i)},\nabla_A\phi^{(i)} .\]

Initially, we set $g^{(0)}=g|_{v=0}$, $\Omega^{(0)}=\Omega|_{v=0}$, $\zeta^{(0)}=\zeta|_{v=0}$,
\[
 (b^{(0)})^A(u,v)=b^A(u,0)-\int_0^v4(\Omega^{(0)})^2
 (\zeta^{(0)})^A(u,v')\,\dd v',
\]
$\chih^{(0)}=\chih|_{v=0}$, $\tr\chi^{(0)}=\tr\chi|_{v=0}$, $\chibh^{(0)}=\chibh|_{v=0}$, $\tr\chib^{(0)}=\tr\chib|_{v=0}$, $\omega^{(0)}=\omega|_{v=0}$, $\omegab^{(0)}=\omegab|_{v=0}$, $(e_3\phi)^{(0)}=e_3\phi|_{v=0}$, $(e_4\phi)^{(0)}=e_4\phi|_{v=0}$, $(\nabla_A\phi)^{(0)}=\nabla_A\phi|_{v=0}$.
For all $i$, we always require 
\[\eta^{(i)}_A=(\zeta^{(i)})^B g^{(i)}_{AB}+\nabla_A\log\Omega^{(i)},\ \etab^{(i)}_A=-(\zeta^{(i)})^B g^{(i)}_{AB}+\nabla_A\log\Omega^{(i)}\]

\noindent\textbf{1. Construction of $(\Omega\chih)^{(i+1/2)}$, $\partial_v\phi^{(i+1)}$, and $\nabla_A\phi^{(i+1)}$}

We use \eqref{eq:Ricci_coefficients_propagation_2} to define $(\Omega\chih)^{(i+1/2)}$,
\begin{equation}
    \left\{
    \begin{aligned}
        & \Lie_{\partial_u+b^{(i)}}(\Omega\chih^{(i+1/2)})_{AB}-\frac{1}{2}(\Omega\tr\chib)^{(i)}(\Omega\chih^{(i+1/2)})_{AB}\\
        &\qquad\qquad -(\Omega\chibh^{(i)})_A^{\quad C}(\Omega\chih^{(i+1/2)})_{CB}-(\Omega\chibh^{(i)})_B^{\quad C}(\Omega\chih^{(i+1/2)})_{CA} \\
        &\qquad =(\Omega^{(i)})^2\left(\left(\nabla^{(i)}{\hat\otimes}\eta^{(i)}+\eta^{(i)}{\hat\otimes}\eta^{(i)}+\frac{1}{2}\nabla\phi^{(i)}\hat\otimes\nabla\phi^{(i)}\right)_{AB}-\frac{1}{2}(\tr\chi)^{(i)}{\chibh^{(i)}}_{AB}\right),\\
        &{\Omega\chih^{(i+1/2)}}_{AB}(-1,v)=\frac{1}{2}\left(\Omega\chih_{AC}g^{CD}{g^{(i)}}_{DB}+\Omega\chih_{BC}g^{CD}{g^{(i)}}_{DA}\right)(-1,v).
    \end{aligned}
    \right.
\end{equation}
Using \eqref{eq:ese-wave-eq}, we define $\partial_v\phi^{(i+1)}$ by
\begin{equation}
    \left\{\begin{aligned}
        &(\Omega e_3)^{(i)}(\partial_v\phi^{(i+1)})+\frac{1}{2}(\Omega\tr\chib)^{(i)}(\partial_v\phi^{(i+1)})\\
        &\qquad =(\Omega^{(i)})^2\Delta^{(i)}\phi^{(i)}-\frac{1}{2}(\Omega\tr\chi)^{(i)}(\Omega e_3)^{(i)}\phi^{(i)}+2(\Omega^{(i)})^2\eta^{(i)}\nabla\phi^{(i)},\\
        &\partial_v\phi^{(i+1)}(-1,v)=\partial_v\phi(-1,v).
    \end{aligned}\right.
\end{equation}
It follows that
\[\nabla_A\phi^{(i+1)}(u,v)=\nabla_A\phi(u,0)+\partial_A\left(\int_0^v\partial_v\phi^{(i+1)}\right).\]

\noindent\textbf{2. Construction of $(\Omega\omegab)^{(i+1/2)}$}

We use \eqref{eq:Ricci_coefficients_propagation_3} to define $(\Omega\omegab)^{(i+1/2)}$,
\begin{equation}
    \left\{
    \begin{aligned}
        &4\partial_v(\Omega\omegab)^{(i+1/2)}=\Omega\chih^{(i+1/2)}_{AB}\Omega\chibh^{(i)}_{CD}(g^{(i)})^{AC}(g^{(i)})^{BD}+\frac{1}{2}(\Omega\tr\chi)^{(i)}(\Omega\tr\chib)^{(i)}\\
        &\qquad-4(\Omega^{(i)})^2\eta^{(i)}\cdot\etab^{(i)}+(\partial_u+b^{(i)})(\Omega\tr\chi)^{(i)}-2(\Omega^{(i)})^2\dv^{(i)}\eta^{(i)}\\
        &\qquad+(\partial_v\phi)^{(i+1)}(\Omega e_3\phi)^{(i)},\\
        &(\Omega\omegab)^{(i+1/2)}(u,0)=\Omega\omegab(u,0).
    \end{aligned}\right.
\end{equation}

\noindent\textbf{3. Construction of $\Omega^{(i+1)}$, $\nabla_A\log\Omega^{(i+1)}$, and $(\Omega\omega)^{(i+1)}$}

With $(\Omega\omegab)^{(i+1/2)}$, we can integrate it to obtain $\Omega^{(i+1)}$
\begin{equation} 
    \left\{
    \begin{aligned}
        &\left(\partial_u+b^{(i)}\right)\log \Omega^{(i+1)}=-2(\Omega\omegab)^{(i+1/2)},\\
        &\log\Omega^{(i+1)}(-1,v)=\log\Omega(-1,v),
    \end{aligned}
    \right.
\end{equation}
\begin{equation}
    \nabla_A\log\Omega^{(i+1)}=\partial_A \log\Omega^{(i+1)}.
\end{equation}
We construct $(\Omega\omega)^{(i+1)}$ by integration instead of differentiation, 
\begin{equation} 
    \left\{\begin{aligned}
        4\left(\partial_u+b^{(i)}\right)&(\Omega\omega)^{(i+1)}=\Omega\chih^{(i+1/2)}_{AB}\Omega\chibh^{(i)}_{CD}(g^{(i)})^{AC}(g^{(i)})^{BD}+\frac{1}{2}(\Omega\tr\chi)^{(i)}(\Omega\tr\chib)^{(i)}\\
        &-4(\Omega^{(i)})^2\eta^{(i)}\cdot\etab^{(i)}+(\partial_u+b^{(i)})(\Omega\tr\chi)^{(i)}-2(\Omega^{(i)})^2\dv^{(i)}\eta^{(i)}\\
        &+(\partial_v\phi)^{(i+1)}(\Omega e_3\phi)^{(i)}-8(\Omega^{(i)})^2(\zeta^{(i)})^A\nabla_A\log\Omega^{(i+1)},\\
        (\Omega\omega)^{(i+1)}&(-1,v)=\Omega\omega(-1,v).
    \end{aligned}\right.
\end{equation}
We note that the construction of $(\Omega\omega)^{(i+1)}$ is equivalent to $(\Omega\omega)^{(i+1)}=-\frac{1}{2}\partial_v\log\Omega^{(i+1)}$, because 
\begin{equation}
   \begin{aligned}
     4(\Omega e_3)^{(i)}(\Omega\omega)^{(i+1)}=&-2\partial_v(\Omega e_3)^{(i)}\log\Omega^{(i+1)}+2\Lie_v b^{(i)}\nabla\log\Omega^{(i+1)}\\
     =&4\partial_v(\Omega\omegab)^{(i+1/2)}-8(\Omega^{(i)})^2(\zeta^{(i)})^A\nabla_A\log\Omega^{(i+1)}.
   \end{aligned}
\end{equation}

\noindent\textbf{4. Construction of $\zeta^{(i+1)}$, $b^{(i+1)}$, and $(\Omega\omegab)^{(i+1)}$}

From \eqref{eq:Ricci_coefficients_propagation_4}, we define $\zeta^{(i+1)}$,
\begin{equation} 
    \left\{\begin{aligned}
        &\Lie_v(\zeta^{(i+1)})^A=-2(\Omega\tr\chi)^{(i)}(\zeta^{(i)})^A-2(g^{(i)})^{AC}\left((\Omega\chih)^{(i+1/2)}\right)_{CB}(\zeta^{(i)})^B\\
        &\qquad+2(g^{(i)})^{AB}\partial_B(\Omega\omega)^{(i+1)} +(\dv)^{(i)}(\Omega\chih)^{(i+1/2)}_B(g^{(i)})^{BA} \\
        &\qquad-\frac{1}{2}(g^{(i)})^{AB}\partial_B(\Omega\tr\chi)^{(i)}+(\Omega\tr\chi)^{(i)}(g^{(i)})^{AB}\partial_B\log\Omega^{(i+1)}\\
        &\qquad-(\partial_v\phi)^{(i+1)}\partial_B\phi^{(i)}(g^{(i)})^{AB},\\
        &(\zeta^{(i+1)})^A(u,0)=\zeta^A(u,0).
    \end{aligned}\right.
\end{equation}
We remark that it is necessary to treat $\zeta^{(i+1)}$ as a vector field instead of a 1-form in order to define $b^{(i+1)}$,
\begin{equation} 
    \left\{\begin{aligned}
        &\Lie_v (b^{(i+1)})^A=-4(\Omega^{(i+1)})^2(\zeta^{(i+1)})^A,\\
        &(b^{(i+1)})^A(u,0)=b^A(u,0).
    \end{aligned}\right.
\end{equation}

The incoming lapse coefficient must then use the updated shift:
\begin{equation}
    (\Omega\omegab)^{(i+1)}=(\Omega\omegab)^{(i+1/2)}
    -\frac12\bigl(b^{(i+1)}-b^{(i)}\bigr)\cdot\nabla\log\Omega^{(i+1)}.
\end{equation}
The numerical implementation applies the angular projection to this correction
and restores the prescribed traces after projection and relaxation.

\noindent\textbf{5. Construction of $g^{(i+1)}$ and $(\Omega\chi)^{(i+1)}$}

It remains to define $g^{(i+1)}$. We consider the following $(g,\tr\chi,\chih)$ system,
\begin{equation}
    \left\{
    \begin{aligned}
        &\Lie_v(g^{(i+1)})_{AB}=(\Omega\tr\chi)^{(i+1)}(g^{(i+1)})_{AB}+2{(\Omega\chih)^{(i+1)}}_{AB},\\
        &\partial_v\left(\Omega\tr\chi\right)^{(i+1)}=-\frac{1}{2}\left((\Omega\tr\chi)^{(i+1)}\right)^2-4(\Omega\omega)^{(i+1)}(\Omega\tr\chi)^{(i+1)}\\
        &\qquad\qquad- {(\Omega\chih)^{(i+1)}}_{AB}{(\Omega\chih)^{(i+1)}}_{CD}{g^{(i+1)}}^{AC}{g^{(i+1)}}^{BD}-(\partial_v\phi^{(i+1)})^2,\\
        &(\Omega\chih)^{(i+1)}_{AB}=\frac{1}{2}\left((\Omega\chih)^{(i+1/2)}_{AC}{(g^{(i)})}^{CD}{g^{(i+1)}}_{DB}
        +(\Omega\chih)^{(i+1/2)}_{BC}{(g^{(i)})}^{CD}{g^{(i+1)}}_{DA}\right),\\
        &(g^{(i+1)})_{AB}(u,0)=g_{AB}(u,0),\ \ (\Omega\tr\chi)^{(i+1)}(u,0)=\Omega\tr\chi(u,0).
    \end{aligned}
    \right.
\end{equation}

\noindent\textbf{6. Construction of $(\Omega\chib)^{(i+1)}$ and $(\Omega e_3\phi)^{(i+1)}$}

There are two equivalent ways to define $(\Omega\chib)^{(i+1)}$. The first one is using $\Omega\chib_{AB}=\frac{1}{2}\Lie_{\Omega e_3}g_{AB}$. We use the second way, which is using equation
\begin{equation}
    \left\{\begin{aligned}
        &\Lie_{v}(\Omega\chib)^{(i+1)}_{AB}= \Lie_{\partial_u+b^{(i+1)}}(\Omega\chi)^{(i+1)}_{AB}-2(\Omega^{(i+1)})^2\left(\nabla^{(i+1)}_A\zeta^{(i+1)}_B+\nabla^{(i+1)}_B\zeta^{(i+1)}_A\right)\\
        &\qquad\qquad-4(\Omega^{(i+1)})^2\left(\zeta^{(i+1)}_A\nabla_B\log\Omega^{(i+1)}+\zeta^{(i+1)}_B\nabla_A\log\Omega^{(i+1)}\right),\\
        &(\Omega\chib)^{(i+1)}_{AB}(u,0)= \Omega\chib_{AB}(u,0).
    \end{aligned}\right.
\end{equation}
We remark that
\begin{equation}
    \begin{aligned}
        &\Omega\nabla_4(\Omega\chib)_{AB}-\Omega\nabla_3(\Omega\chi)_{AB}\\
        =&\Lie_{v}(\Omega\chib)_{AB}-(\Omega\chi)_{A}^{\ C}(\Omega\chib)_{CB}-(\Omega\chi)_{B}^{\ C}(\Omega\chib)_{CA}\\
        &-\Lie_{\partial_u+b}(\Omega\chi)_{AB}+(\Omega\chib)_{A}^{\ C}(\Omega\chi)_{CB}+(\Omega\chib)_{B}^{\ C}(\Omega\chi)_{AC}\\
        =&\Lie_{v}(\Omega\chib)_{AB}-\Lie_{\partial_u+b}(\Omega\chi)_{AB}.
    \end{aligned}
\end{equation}
Theoretically, $(\Omega\chib)_{AB}^{(i+1)}=\frac{1}{2}\Lie_{\partial_u+b^{(i+1)}}g^{(i+1)}_{AB}.$ Similarly, we construct $(\Omega e_3\phi)^{(i+1)}$ by
\begin{equation}
    \left\{\begin{aligned}
        &\partial_v(\Omega e_3\phi)^{(i+1)}=(\partial_u+b^{(i+1)})\partial_v\phi^{(i+1)}-4(\Omega^{(i+1)})^2(\zeta^{(i+1)})^A\nabla_A\phi^{(i+1)},\\
        &(\Omega e_3\phi)^{(i+1)}(u,0)=\Omega e_3\phi(u,0).
    \end{aligned}\right.
\end{equation}

For Einstein vacuum equation
\[\Rc_{\mu\nu}=0,\]
removing all $\phi$-terms of ESE gives the iterative construction directly, so we omit the repeation of construction.

\subsection{Curvature residues}
Computing curvature usually involves second order derivatives, which would heavily reduce the numerical precision. 
Especially, $\lim_{v\rightarrow 0}R_{4A4B}(-1,v)$ can be infinite for short pulse perturbation and $\Rc_{44}$ is hard to compute in this way, even though it is theoretically zero in our construction.
See \cite{an-25} for the example of short pulse with infinite $R_{4A4B}$.
We use intrinsic equations of Lorentzian manifold to compute the Ricci curvature:
\[\Omega^2\Rc_{33}=-\Omega\nabla_3(\Omega\tr\chib)-4\Omega\omegab\Omega\tr\chib-\frac{1}{2}(\Omega\tr\chib)^2-|\Omega\chibh|^2,\]
\[\Omega\Rc_{3A}=\Omega\nabla_3\zeta+\frac{3}{2}\Omega\tr\chib\zeta+\Omega\chibh\cdot\zeta+2\nabla(\Omega\omegab)+\dv(\Omega\chibh)-\frac{1}{2}\nabla(\Omega\tr\chib)+\Omega\tr\chib\nabla\log\Omega,\]
\[\Omega\Rc_{4A}=-\Omega\nabla_4\zeta-\frac{3}{2}\Omega\tr\chi\zeta-\Omega\chih\cdot\zeta+2\nabla(\Omega\omega)+\dv(\Omega\chih)-\frac{1}{2}\nabla(\Omega\tr\chi)+\Omega\tr\chi\nabla\log\Omega,\]
\[\Omega^2\widehat{\Rc}_{AB}=\Omega\nabla_3(\Omega\chih)+\frac{1}{2}\Omega\tr\chib\,\Omega\chih-\Omega^2\left(\nabla\hat\otimes\eta+\eta\hat\otimes\eta\right)+\frac{1}{2}\Omega\tr\chi\,\Omega\chibh,\]
\[ \Omega^2g^{AB}\Rc_{AB}=\Omega\nabla_3(\Omega\tr\chi)+\Omega\tr\chi\Omega\tr\chib-2\Omega^2\dv\eta-2\Omega^2|\eta|^2+2\Omega^2K,\]
\[\Omega^2\Rc_{34}=4\Omega\nabla_4(\Omega\omegab)-\Omega\chih\cdot\Omega\chibh-\frac{1}{2}\Omega\tr\chi\Omega\tr\chib+4\Omega^2\eta\cdot\etab -\Omega\nabla_3(\Omega\tr\chi)+2\Omega^2\dv\eta.\]
For a fresh check of the outgoing constraint we also use
\[
 \Omega^2\Rc_{44}=-\partial_v(\Omega\tr\chi)
 -4(\Omega\omega)(\Omega\tr\chi)
 -\tfrac12(\Omega\tr\chi)^2-|\Omega\chih|^2.
\]
The coefficient $4$ includes the product-rule contribution from weighting
$\tr\chi$ by $\Omega$, since $\partial_v\log\Omega=-2\Omega\omega$.
Thus no outgoing null derivative of $\chih$ is required.  These formulas use
first null derivatives of the stored connection variables; the intrinsic
Gauss curvature $K$ uses the angularly smooth section metric.  The numerical
metric--connection relations are checked separately.  Endpoint values at
singular coordinate Jacobians are excluded from the reported residuals.

\subsection{Legendre--Gauss--Lobatto spectral elements}

In the experiments, we will add non-spherical symmetric perturbation to the Einstein scalar-field or vacuum systems. To make the metric less regular, the perturbation is required to be of size $O(v^\delta)$.
To deal with the non-smoothness, the characteristic coordinates are first replaced by coordinates in which the
corner profiles are smoother.  In particular, we use
\[
    \tau=-\log(-u),\qquad
    s=\left(\frac{v}{v_{\max}}\right)^\delta ,
    \qquad
    u=-e^{-\tau},\qquad
    v=v_{\max}s^{1/\delta}.
\]
Thus a profile proportional to $v^\delta$ is linear in $s$.  This change of
variables does not regularize the physical solution; it only represents its
fractional behavior by a smooth function of the computational coordinate.
Physical derivatives and integrals are recovered by
\[
    \partial_u=\frac{1}{-u}\partial_\tau,\qquad
    \partial_v=\left(\frac{dv}{ds}\right)^{-1}\partial_s,\qquad
    \int_0^v F(v')\,dv'
    =\int_0^s F(v(s'))\frac{dv}{ds'}\,ds'.
\]

We divide each computational coordinate interval into elements.  On an
element of degree $p$, let $\{\xi_j\}_{j=0}^p$ be the
Legendre--Gauss--Lobatto nodes, characterized by
\[
    (1-\xi_j^2)P_p'(\xi_j)=0,
\]
including the two endpoints.  A component $F$ is represented by its Lagrange
interpolant
\[
    I_pF(\xi)=\sum_{j=0}^p F(\xi_j)\ell_j(\xi).
\]
The element differentiation and indefinite-integration matrices are
\[
    D_{ij}=\ell_j'(\xi_i),\qquad
    Q_{ij}=\int_{-1}^{\xi_i}\ell_j(\xi)\,d\xi.
\]
Consequently, interpolation of intermediate Runge--Kutta stages,
differentiation, and cumulative integration are all obtained from the same
element polynomial.  Neighboring elements share their endpoint value, while
the two derivative traces at an interior interface are combined into a
single interface derivative.  This composite construction permits local
refinement in either characteristic direction without replacing the
underlying transport hierarchy.

\subsection{Tensor spherical-harmonic Galerkin discretization}
The angular variables are represented on a quasi-uniform, pole-free point set
on $\mathbb S^2$.  Geometric vectors and tensors are stored as ambient
Cartesian tensors and projected to the tangent bundle.  This avoids the
coordinate singularities of a polar chart.  For a scalar field, we use the
real spherical-harmonic expansion
\[
    f_L(\vartheta)
    =\sum_{\ell=0}^{L}\sum_{m=-\ell}^{\ell}
      f_{\ell m}Y_{\ell m}(\vartheta).
\]
The coefficients are obtained by an overdetermined least-squares analysis of
the nodal values, and angular derivatives are computed by differentiating the
harmonic basis.

Vectors and symmetric two-tensors are not expanded componentwise as
unrelated scalars.  We use the electric and magnetic vector harmonics
\[
    Y_A^{E,\ell m}\sim\nabla_A Y_{\ell m},\qquad
    Y_A^{B,\ell m}\sim\epsilon_A{}^B\nabla_B Y_{\ell m},
\]
and the trace, electric, and magnetic tensor harmonics
\[
    \gamma_{AB}Y_{\ell m},\qquad
    Y_{AB}^{E,\ell m}\sim
       \left(\nabla_A\nabla_B Y_{\ell m}\right)^{\mathrm{TF}},
    \qquad
    Y_{AB}^{B,\ell m}\sim
       \epsilon_{(A}{}^C\nabla_{B)}\nabla_CY_{\ell m}.
\]
These typed spaces preserve tangency, symmetry, and the tensorial
transformation law.  Trace-free quantities are projected with respect to the
current section metric,
\[
    T_{AB}^{\mathrm{TF}}
    =T_{AB}-\frac12(\tr_\gamma T)\gamma_{AB},
\]
rather than with respect to a fixed background metric.

Let $\mathcal V_L$ denote the retained harmonic space and let
$\mathcal V_W$, with $W>L$, be a work space used to form angular products.
For a semidiscrete unknown $U_L\in\mathcal V_L$, every nonlinear right-hand
side is defined by
\[
    \partial U_L=\Pi_L F(U_L),
\]
where $F(U_L)$ is evaluated in the work space and $\Pi_L$ is the appropriate
scalar, vector, or tensor Galerkin projection.  The retained/work separation
reduces angular aliasing while keeping the evolved state in one declared
finite-dimensional geometric space.
\section{Results for Einstein Vacuum Equations}

We test the vacuum iteration in five regimes of increasing geometric and
numerical difficulty.  The first three experiments admit explicit reference
metrics and therefore measure the error directly by comparing the computed
section metric with the exact one.  The last two experiments evolve genuinely
nonspherical characteristic data, for which no explicit spacetime is
available; there we report a positive norm of the first-order Ricci residual
on a protected interior region, together with metric--connection consistency
checks.  The residual is evaluated from the stored weighted connection
variables and fresh derivatives, with the audit grid specified for each case.  This avoids
second null derivatives of the metric and the potentially singular derivative
$\nabla_4\chih$ at the initial corner; intrinsic angular curvature uses the
smooth sphere data.  These diagnostics assess the numerical equations and do
not furnish a certified metric-error bound.

\subsection{Regular Schwarzschild and Kerr benchmarks}

We begin with the Schwarzschild and subextremal Kerr families.  In
Schwarzschild coordinates, the metric of mass $M>0$ is
\begin{equation}\label{eq:results-schwarzschild-static}
  \bg_{\mathrm{Sch}}
  =-\left(1-\frac{2M}{r}\right)\dd t^2
   +\left(1-\frac{2M}{r}\right)^{-1}\dd r^2
   +r^2\left(\dd\vartheta^2+\sin^2\vartheta\,\dd\varphi^2\right).
\end{equation}
Let
\begin{equation}
  r_*=r+2M\log\left(\frac{r}{2M}-1\right)+C,
  \qquad t=u+v,\qquad r_*=v-u,
\end{equation}
where the additive constant $C$ fixes the reference sphere.  Then
\eqref{eq:results-schwarzschild-static} becomes
\begin{equation}\label{eq:results-schwarzschild-double-null}
  \bg_{\mathrm{Sch}}
  =-4\Omega^2\dd u\dd v+\slg_{AB}\dd\theta^A\dd\theta^B,
  \qquad
  \Omega^2=1-\frac{2M}{r},\qquad
  \slg=r^2\roundg,\qquad b=0,
\end{equation}
with $r=r(v-u)$.  Thus the Schwarzschild benchmark is already in the
double-null gauge used by the iteration.

For Kerr, write $a$ for the angular momentum per unit mass and set
\begin{equation}
  \rho^2=r^2+a^2\cos^2\vartheta,\qquad
  \Delta=r^2-2Mr+a^2.
\end{equation}
The Boyer--Lindquist expression is
\begin{align}\label{eq:results-kerr-bl}
  \bg_{\mathrm K}={}&
  -\left(1-\frac{2Mr}{\rho^2}\right)\dd t^2
  -\frac{4Mar\sin^2\vartheta}{\rho^2}\dd t\dd\varphi
  +\frac{\rho^2}{\Delta}\dd r^2+\rho^2\dd\vartheta^2 \notag\\
  &+\frac{\sin^2\vartheta}{\rho^2}
  \left((r^2+a^2)^2-a^2\Delta\sin^2\vartheta\right)\dd\varphi^2.
\end{align}
To put \eqref{eq:results-kerr-bl} into double-null form, we use the optical
coordinates $(u,v,\vartheta_*,\varphi_*)$ defined by
\begin{equation}\label{eq:results-kerr-coordinate-map}
  t=u+v,\qquad s=u-v,\qquad
  r=r(s,\vartheta_*),\qquad
  \vartheta=\vartheta(s,\vartheta_*),\qquad
  \varphi=\varphi_*+h(s,\vartheta_*).
\end{equation}
The functions $r$, $\vartheta$, and $h$ are obtained from the Kerr optical
map, with $r=4M$, $\vartheta=\vartheta_*$, and $h=0$ on the reference
two-sphere.  If $\widetilde g_{\alpha\beta}$ denotes the pullback of
\eqref{eq:results-kerr-bl} under \eqref{eq:results-kerr-coordinate-map}, the
optical equations give
\begin{equation}
  \widetilde g_{vv}=0,\qquad \widetilde g_{vA}=0,\qquad
  \widetilde g_{uu}=\slg_{AB}b^Ab^B.
\end{equation}
Consequently the pulled-back metric is exactly
\begin{align}\label{eq:results-kerr-double-null}
  \bg_{\mathrm K}
  &=-4\Omega^2\dd u\dd v
   +\slg_{AB}(\dd\theta^A-b^A\dd u)
                  (\dd\theta^B-b^B\dd u), \notag\\
  \Omega^2&=-\frac12\widetilde g_{uv},\qquad
  \slg_{AB}=\widetilde g_{AB},\qquad
  b^A=-\slg^{AB}\widetilde g_{uB}.
\end{align}

All runs use $M=1$.  We take $a=0,0.3,0.7,0.9$ and solve on the short
rectangle
\begin{equation}
  -1\leq u\leq-\frac12,\qquad 0\leq v\leq\frac12,
\end{equation}
and on the long rectangle
\begin{equation}
  -1\leq u\leq0,\qquad 0\leq v\leq1.
\end{equation}
The coordinate refinements contain $17^2$, $33^2$, and $65^2$ nodes.  The
Schwarzschild runs use 50 spherical points with retained degree five, while
the finest Kerr results below use 86 spherical points with retained degree
seven.  Eight complete Picard sweeps are used in every case.

Because the exact metrics are known, no curvature differentiation is needed
to define the primary error.  At every angular and double-null grid point
$q=(\theta,u_i,v_j)$, set
\begin{equation}\label{eq:results-exp1-metric-error}
  E_g(q)=\left\lVert
  \slg_{\mathrm{num}}(q)-\slg_{\mathrm{exact}}(q)
  \right\rVert_{\mathrm F},
\end{equation}
where $\norm{\cdot}_{\mathrm F}$ is the Frobenius norm in the pole-free
ambient representation of tangent two-tensors.  Table
\ref{tab:results-exp1-metric-error} gives the median and arithmetic mean of
$E_g$ over all points of each $65\times65$ run.  Schwarzschild and the
zero-spin Kerr control agree with the explicit metric essentially to
floating-point accuracy.  For $a\neq0$, the larger error is associated with
the numerical angular reconstruction of the Kerr optical map; it remains of
order $10^{-6}$ in the mean for all rotating cases considered here.

\begin{table}[htbp]
  \centering
  \caption{Final section-metric error \eqref{eq:results-exp1-metric-error} on
  the finest coordinate grid.}
  \label{tab:results-exp1-metric-error}
  \begin{tabular}{llrrr}
    \toprule
    Family & Domain & $\operatorname{median} E_g$
      & $\operatorname{mean} E_g$ & $\max E_g$ \\
    \midrule
    Schwarzschild & short & $2.168\times10^{-14}$ & $2.387\times10^{-14}$ & $1.087\times10^{-13}$ \\
Schwarzschild & long & $3.782\times10^{-13}$ & $3.686\times10^{-13}$ & $9.207\times10^{-13}$ \\
Kerr, $a=0.0$ & short & $1.264\times10^{-14}$ & $1.313\times10^{-14}$ & $4.252\times10^{-14}$ \\
Kerr, $a=0.0$ & long & $3.445\times10^{-13}$ & $3.387\times10^{-13}$ & $7.693\times10^{-13}$ \\
Kerr, $a=0.3$ & short & $2.892\times10^{-7}$ & $5.727\times10^{-7}$ & $8.037\times10^{-6}$ \\
Kerr, $a=0.3$ & long & $1.120\times10^{-6}$ & $2.356\times10^{-6}$ & $3.918\times10^{-5}$ \\
Kerr, $a=0.7$ & short & $4.442\times10^{-7}$ & $9.360\times10^{-7}$ & $1.792\times10^{-5}$ \\
Kerr, $a=0.7$ & long & $1.888\times10^{-6}$ & $4.384\times10^{-6}$ & $1.030\times10^{-4}$ \\
Kerr, $a=0.9$ & short & $5.247\times10^{-7}$ & $9.864\times10^{-7}$ & $1.753\times10^{-5}$ \\
Kerr, $a=0.9$ & long & $2.446\times10^{-6}$ & $5.373\times10^{-6}$ & $1.022\times10^{-4}$ \\
    \bottomrule
  \end{tabular}
\end{table}

\subsection{Schwarzschild horizon in static and Kruskal coordinates}

The second experiment keeps the Schwarzschild geometry
\eqref{eq:results-schwarzschild-static} fixed and changes only the coordinate
system.  It separates loss of numerical conditioning caused by a degenerating
chart from the regular geometry of the event horizon.  In the static
double-null chart \eqref{eq:results-schwarzschild-double-null}, the additive
constant in $r_*$ is chosen so that
\begin{equation}
  r=2M(1+\epsilon)
  \quad\text{at}\quad v-u=\frac12 .
\end{equation}
The four offsets
\begin{equation}
  \epsilon=10^{-1},\ 10^{-2},\ 10^{-4},\ 10^{-6}
\end{equation}
therefore give
\begin{equation}
  \min\Omega^2=\frac{\epsilon}{1+\epsilon}.
\end{equation}
Although every one of these rectangles remains in the exterior $r>2M$, the
last case places the closest numerical sphere only $2M\epsilon=2\times10^{-6}$
from the horizon when $M=1$.  The lapse consequently becomes small and the
static coordinate representation becomes increasingly ill-conditioned.

For the horizon-crossing test, define shifted Kruskal coordinates
\begin{equation}
  U=u+\frac34,\qquad V=v+1.
\end{equation}
The relation between $r$ and $(U,V)$ and the corresponding double-null metric
are
\begin{equation}\label{eq:results-kruskal-double-null}
  UV=\left(1-\frac{r}{2M}\right)e^{r/(2M)},\qquad
  \bg_{\mathrm{Sch}}
  =-\frac{32M^3}{r}e^{-r/(2M)}\dd U\dd V+r^2\roundg,
\end{equation}
or, in the notation of this paper,
\begin{equation}
  \Omega^2=\frac{8M^3}{r}e^{-r/(2M)},\qquad
  \slg=r^2\roundg,\qquad b=0.
\end{equation}
Unlike the static lapse, this $\Omega$ is positive and regular at $r=2M$.
The principal Lambert branch gives the explicit inversion
\begin{equation}
  r=2M\left[1+
  W_0\left(-\frac{UV}{e}\right)\right].
\end{equation}

All five cases are solved on the same numerical rectangle
\begin{equation}\label{eq:results-exp2-domain}
  -1\leq u\leq-\frac12,\qquad 0\leq v\leq\frac12,
\end{equation}
using $17^2$, $33^2$, and $65^2$ coordinate grids, 50 spherical points,
retained angular degree five, and eight Picard sweeps.  For the Kruskal case,
\eqref{eq:results-exp2-domain} corresponds to
\begin{equation}
  -\frac14\leq U\leq\frac14,\qquad
  1\leq V\leq\frac32.
\end{equation}
It crosses the future event horizon $U=0$ and covers
$1.67549\leq r\leq2.24420$ for $M=1$.

We again use the direct metric error $E_g$ defined in
\eqref{eq:results-exp1-metric-error}.  Table
\ref{tab:results-exp2-metric-error} reports its median and arithmetic mean on
the finest grid.  The static-coordinate metric itself is reproduced to
roundoff even as $\Omega^2$ approaches zero.  The Kruskal calculation also
remains highly accurate across the horizon, with a mean section-metric error
of $1.12\times10^{-12}$.  The supplementary curvature checks use first null derivatives of the
weighted connection variables, with the metric--connection consistency
reported separately.
Derivative-based curvature diagnostics are more sensitive to the static lapse
degeneration; this is a coordinate-conditioning effect and does not indicate
a curvature singularity at $r=2M$.

\begin{table}[htbp]
  \centering
  \caption{Final section-metric error for the Schwarzschild horizon tests on
  the $65\times65$ coordinate grid.}
  \label{tab:results-exp2-metric-error}
  \begin{tabular}{lrrr}
    \toprule
    Coordinates & $\operatorname{median} E_g$
      & $\operatorname{mean} E_g$ & $\max E_g$ \\
    \midrule
    Static, $\epsilon=10^{-1}$ & $2.627\times10^{-15}$ & $2.800\times10^{-15}$ & $1.358\times10^{-14}$ \\
Static, $\epsilon=10^{-2}$ & $2.473\times10^{-15}$ & $2.749\times10^{-15}$ & $1.435\times10^{-14}$ \\
Static, $\epsilon=10^{-4}$ & $1.858\times10^{-15}$ & $1.960\times10^{-15}$ & $8.129\times10^{-15}$ \\
Static, $\epsilon=10^{-6}$ & $1.776\times10^{-15}$ & $1.873\times10^{-15}$ & $7.501\times10^{-15}$ \\
Kruskal crossing & $1.134\times10^{-13}$ & $1.122\times10^{-12}$ & $2.923\times10^{-11}$ \\
    \bottomrule
  \end{tabular}
\end{table}

\subsection{Schwarzschild interior and approach to \texorpdfstring{\(r=0\)}{r=0}}

The third experiment uses the Schwarzschild solution in the regular Kruskal
form \eqref{eq:results-kruskal-double-null}, now entirely inside the event
horizon.  We shift the Kruskal coordinates by
\begin{equation}
  U=u+2,\qquad V=v+\frac15,
\end{equation}
and use the same relation
\begin{equation}
  UV=\left(1-\frac{r}{2M}\right)e^{r/(2M)}.
\end{equation}
The singular boundary $r=0$ corresponds to $UV=1$.  To approach it without
including it, let
\begin{equation}
  x_\epsilon=(1-\epsilon)e^\epsilon,\qquad
  v_0(u)=\frac{x_\epsilon}{u+2}-\frac15,
\end{equation}
and introduce a fixed computational coordinate $\xi$ through
\begin{equation}\label{eq:results-exp3-map}
  v=\xi v_0(u).
\end{equation}
The numerical domain is therefore the curved characteristic region
\begin{equation}\label{eq:results-exp3-domain}
  -1\leq u\leq-\frac12,\qquad
  0\leq\xi\leq1,\qquad
  0\leq v\leq v_0(u).
\end{equation}
On its future boundary $\xi=1$ one has
\begin{equation}
  UV=x_\epsilon,\qquad r=2M\epsilon.
\end{equation}
We take
\begin{equation}
  \epsilon=2^{-1},2^{-2},\ldots,2^{-6}.
\end{equation}
For $M=1$, the six future boundaries thus lie at
$r=1,1/2,1/4,1/8,1/16,1/32$.  The opposite characteristic face has
$r\leq1.84064$, so every case lies strictly within the Schwarzschild event
horizon.

The map \eqref{eq:results-exp3-map} is discretized with tensor-product
Chebyshev--Lobatto grids containing $17^2$, $33^2$, and $49^2$ points.  The
physical derivatives include both Jacobian terms generated by the
$u$-dependence of $v_0$.  Each case receives at most 20 Picard sweeps.  If an
undamped update would make the radius nonpositive, dyadic backtracking chooses
the largest positive, nonincreasing update.  This stabilization prevents an
algebraic breakdown, but it does not by itself establish convergence or
accuracy near $r=0$.

Since the solution is spherically symmetric,
\begin{equation}
  \slg_{\mathrm{num}}=r_{\mathrm{num}}^2\roundg,\qquad
  \slg_{\mathrm{exact}}=r_{\mathrm{exact}}^2\roundg.
\end{equation}
Accordingly, the pointwise metric error
\eqref{eq:results-exp1-metric-error} reduces to
\begin{equation}\label{eq:results-exp3-metric-error}
  E_g=\sqrt{2}\,
  \left|r_{\mathrm{num}}^2-r_{\mathrm{exact}}^2\right|,
\end{equation}
because $\norm{\roundg}_{\mathrm F}=\sqrt2$ in the pole-free ambient
representation.  Table \ref{tab:results-exp3-metric-error} reports the
statistics of \eqref{eq:results-exp3-metric-error} on the $49^2$ grid.

\begin{table}[htbp]
  \centering
  \caption{Final section-metric error in the Schwarzschild-interior
  experiment.  Here $r_{\mathrm{end}}=2M\epsilon$ is the exact radius of the
  future boundary and $M=1$.}
  \label{tab:results-exp3-metric-error}
  \begin{tabular}{ccrrr}
    \toprule
    $\epsilon$ & $r_{\mathrm{end}}$
      & $\operatorname{median}E_g$ & $\operatorname{mean}E_g$ & $\max E_g$ \\
    \midrule
    $2^{-1}$ & $1$ & $2.026\times10^{-7}$ & $5.158\times10^{-7}$ & $2.208\times10^{-6}$ \\
$2^{-2}$ & $1/2$ & $2.810\times10^{-5}$ & $4.039\times10^{-5}$ & $1.364\times10^{-4}$ \\
$2^{-3}$ & $1/4$ & $2.212\times10^{-3}$ & $4.203\times10^{-3}$ & $1.835\times10^{-2}$ \\
$2^{-4}$ & $1/8$ & $6.969\times10^{-3}$ & $1.234\times10^{-2}$ & $6.119\times10^{-2}$ \\
$2^{-5}$ & $1/16$ & $2.020\times10^{-2}$ & $3.743\times10^{-2}$ & $2.851\times10^{-1}$ \\
$2^{-6}$ & $1/32$ & $3.419\times10^{-2}$ & $8.845\times10^{-2}$ & $4.180\times10^{-1}$ \\
    \bottomrule
  \end{tabular}
\end{table}

The first two future boundaries retain small absolute metric error.  Beginning
at $\epsilon=2^{-3}$, however, the iteration does not settle within 20 sweeps
and the error grows rapidly under further approach to the singularity.  In the
closest case, the exact minimum radius is $1/32$, whereas the computed minimum
is $0.08127$; the maximum relative radius error is $16.43$, and the maximum
relative section-metric error is $3.03\times10^2$.  That case also requires a
dyadic relaxation factor as small as $1/32$.  These results demonstrate the
current method's loss of accuracy near $r=0$; the stabilized finite output is
not evidence of convergence at the Schwarzschild singularity.

\subsection{Strong outgoing short pulse and zero control}

The fourth experiment leaves the class of explicit solutions.  At the corner
we prescribe the round section metric, unit lapse, vanishing shift and
torsion, and the Minkowski expansions,
\begin{equation}
  \begin{aligned}
    \slg(-1,0)&=\roundg,& \Omega(-1,0)&=1,&
    b(-1,0)&=0,& \zeta(-1,0)&=0,\\
    (\Omega\tr\chi)(-1,0)&=2,&
    (\Omega\tr\chib)(-1,0)&=-2.&&
  \end{aligned}
\end{equation}
On the outgoing initial hypersurface, a low-band trace-free tensor is formed
from two polynomial hemisphere tensors $X_+$ and $X_-$ and two disjoint
profiles $q_+$ and $q_-$:
\begin{equation}\label{eq:results-exp4-pulse}
  \Omega\chih(-1,v)
  =\mathcal T_{q_+(v)X_++q_-(v)X_-}[\slg(-1,v)].
\end{equation}
Here $q_+$ has the admissible square-root behavior at $v=0$, $q_-$ is a
smooth interior bump, and both are calibrated on $0\leq v\leq1/2$ so that
the hemisphere integral has strength one after division by $3.2$.  Only the
leading part of this pulse is evolved, on the rectangle
\begin{equation}
  -1\leq u\leq-\frac12,\qquad 0\leq v\leq0.005.
\end{equation}
The zero control uses the identical construction and discretization with
$q_+=q_-=0$.

The perturbation in the strong run is not numerically negligible.  Using the
computed metric and area form, we measure
\begin{equation}
  \norm{\chih(-1,0.005)}_{L^2(S,\slg)}
  =\left(\int_{S_{-1,0.005}}
      \abs{\Omega^{-1}\Omega\chih}_{\slg}^{2}\,
      \dd\mu_{\slg}\right)^{1/2}
  =3.75480,
\end{equation}
while the pointwise maximum is $1.26734$.  Thus the good residual reported
below is not a consequence of evolving an effectively zero datum.

We use $17$ nodes in the logarithmic coordinate $\tau=-\log(-u)$ and $33$
nodes in $s=\sqrt{2v}$, distributed over two and four degree-eight elements,
respectively.  The angular discretization has $550$ sphere points, retained
degree $L=10$, work degree $W=20$, and differentiation degree $D=21$.  Both
runs complete six Picard sweeps.  Since no explicit spacetime is available,
the accuracy statistic is the $L^\infty_{u,v}L^2(S)$ norm of the independently
differentiated first-order Ricci residual on the stencil-protected interior.
The audit transfers scalar, vector, and tensor fields in their typed harmonic
spaces through degree 20 onto 562 sphere points, and raises each coordinate
element degree by three. The protected metric--connection component defects
are $1.75\times10^{-9}$ in the incoming direction and
$2.77\times10^{-7}$ in the outgoing direction.

\begin{table}[htbp]
  \centering
  \caption{Iteration update and independent Ricci residual for the strong
  outgoing pulse and its zero control.  The pointwise column is the maximum
  sphere-point residual on the same protected region.}
  \label{tab:results-exp4-residual}
  \begin{tabular}{lrrr}
    \toprule
    Data & final update & $L^\infty_{u,v}L^2(S)$
      & protected pointwise maximum \\
    \midrule
    Strong pulse & $1.688\times10^{-10}$ & $2.617\times10^{-5}$ & $3.297\times10^{-5}$ \\
Zero control & $2.927\times10^{-13}$ & $1.973\times10^{-9}$ & $1.271\times10^{-9}$ \\
    \bottomrule
  \end{tabular}
\end{table}

Every boundary spectral-deferred-correction element is accepted.  For the
strong pulse, the maximum collocation and overgrid boundary defects are
$6.36\times10^{-16}$ and $1.47\times10^{-8}$, respectively.  In the zero
control, moving the residual mask farther from the short-pulse endpoint
reduces the $L^\infty_{u,v}L^2(S)$ residual from $1.973\times10^{-9}$ to
$1.737\times10^{-9}$. The exact Minkowski control has the same numerical
floor, $1.973\times10^{-9}$.  The contrast between this numerical floor and the
strong-pulse value isolates the error associated with evolving the
nonspherical datum.

\subsection{Crossed characteristic shears}

The fifth experiment prescribes nonspherical shears on both initial null
hypersurfaces.  Let
\begin{equation}
  A_1=\operatorname{diag}\left(-\frac12,-\frac12,1\right),\qquad
  A_2=\operatorname{diag}(1,-1,0),\qquad
  f_j(n)=n^{\mathsf T}A_jn,
\end{equation}
and define the explicitly normalized round-sphere tensors
\begin{equation}
  X_1=\frac{\sqrt2}{3}\,\tf\nabla^2 f_1,
  \qquad
  X_2=\frac{1}{2\sqrt2}\,\tf\nabla^2 f_2.
\end{equation}
If $\mathcal T_X[\slg]$ denotes the symmetric transfer of $X$ to $\slg$
followed by removal of its $\slg$-trace, the prescribed data are
\begin{equation}\label{eq:results-exp5-shears}
  \Omega\chih(-1,v)=v^{1/2}\mathcal T_{X_1}[\slg(-1,v)],
  \qquad
  \Omega\chibh(u,0)=(u+1)^{1/2}
     \mathcal T_{X_2}[\slg(u,0)].
\end{equation}
At the corner, $\slg=\roundg$, $\Omega=1$, $\Omega\tr\chi=2$,
$\Omega\tr\chib=-2$, and $\zeta=0$.  The lapse is one on both initial
faces and $b(u,0)=0.1\,\partial_\varphi$; the remaining face data are obtained
from the characteristic constraints.

Although the shears in \eqref{eq:results-exp5-shears} vanish continuously at
the corner, the corresponding initial curvature is singular.  The outgoing
shear equation contains $\nabla_4\chih$ and the incoming shear equation
contains $\nabla_3\chibh$.  Their leading transverse derivatives satisfy
\begin{equation}
  \nabla_4\chih=O(v^{-1/2}),\qquad
  \nabla_3\chibh=O((u+1)^{-1/2}).
\end{equation}
Consequently, both $R_{4A4B}$ on $u=-1$ and $R_{3A3B}$ on $v=0$ are
unbounded and hence infinite at the corner in the limiting sense.  This is a
singularity of the prescribed initial curvature, not a large-amplitude
divergence of the shears themselves.  Indeed, at the opposite endpoints of
the two initial faces,
\begin{equation}
  \norm{\chih(-1,\tfrac12)}_{L^2(S,\slg)}=2.727749,
  \qquad
  \norm{\chibh(-\tfrac12,0)}_{L^2(S,\slg)}=0.785641.
\end{equation}

The solution region is the full rectangle
\begin{equation}
  -1\leq u\leq-\frac12,\qquad 0\leq v\leq\frac12.
\end{equation}
We discretize the square-root coordinates
$t=\sqrt{2(u+1)}$ and $s=\sqrt{2v}$ using six degree-ten elements in each
direction, giving a $61\times61$ coordinate grid.  The angular calculation
uses $300$ points with $L=8$, $W=14$, and $D=15$.  The evolution proceeds in
six successive $v$-slabs.  Their terminal weighted updates are
\begin{equation}
  (1.850,\ 4.897,\ 8.694,\ 4.404,\ 6.984,\ 4.761)\times10^{-9},
\end{equation}
all below the prescribed $10^{-8}$ tolerance.

\begin{table}[htbp]
  \centering
  \caption{Aggregate Ricci residual for the crossed-shear experiment.}
  \label{tab:results-exp5-residual}
  \begin{tabular}{lrrrr}
    \toprule
    Region & Number of sections
      & $\operatorname{median}E_{\mathrm{Ric}}$
      & $\operatorname{mean}E_{\mathrm{Ric}}$
      & $\max E_{\mathrm{Ric}}$ \\
    \midrule
    Protected mask & 144 & $3.960\times10^{-4}$ & $5.914\times10^{-4}$ & $3.432\times10^{-3}$ \\
Open grid & 3481 & $1.287\times10^{-3}$ & $8.947\times10^{0}$ & $1.371\times10^{3}$ \\
    \bottomrule
  \end{tabular}
\end{table}

Let $E_{\mathrm{Ric}}(u,v)$ denote the sum of the sectionwise $L^2(S_{u,v})$
norms of the six weighted Ricci quantities listed in the curvature-residue
subsection.  Table \ref{tab:results-exp5-residual} compares this aggregate
residual on the protected mask with the same statistic on the open
coordinate grid, excluding the four outer endpoint lines.

The protected mask is constructed separately in the two characteristic
coordinates.  In each coordinate it removes the complete first and last
spectral elements and three LGL nodes on each side of every remaining
element interface.  The two-dimensional mask is the tensor product of these
one-dimensional masks, leaving $12$ retained nodes in each direction and
hence $144$ sections.  Its purpose is to omit locations at which the
independent derivative reconstruction encounters the square-root endpoint,
a terminal one-sided boundary, or an element interface; the open-grid row retains the nearby endpoint layers and element interfaces,
but excludes the endpoint lines themselves.  The minimum section-metric eigenvalue is
$2.938\times10^{-2}$, and all immutable characteristic traces are restored
to roundoff, with maximum incoming $\Omega\tr\chi$ mismatch
$1.78\times10^{-15}$.  On the protected mask the metric--connection
defects $\partial_u g+\Lie_b g-2\Omega\chib$ and
$\partial_v g-2\Omega\chi$ have maximum absolute ambient components
$3.10\times10^{-4}$ and $1.85\times10^{-5}$, respectively.

\begin{figure}[htbp]
  \centering
  \includegraphics[width=0.8\textwidth]{
    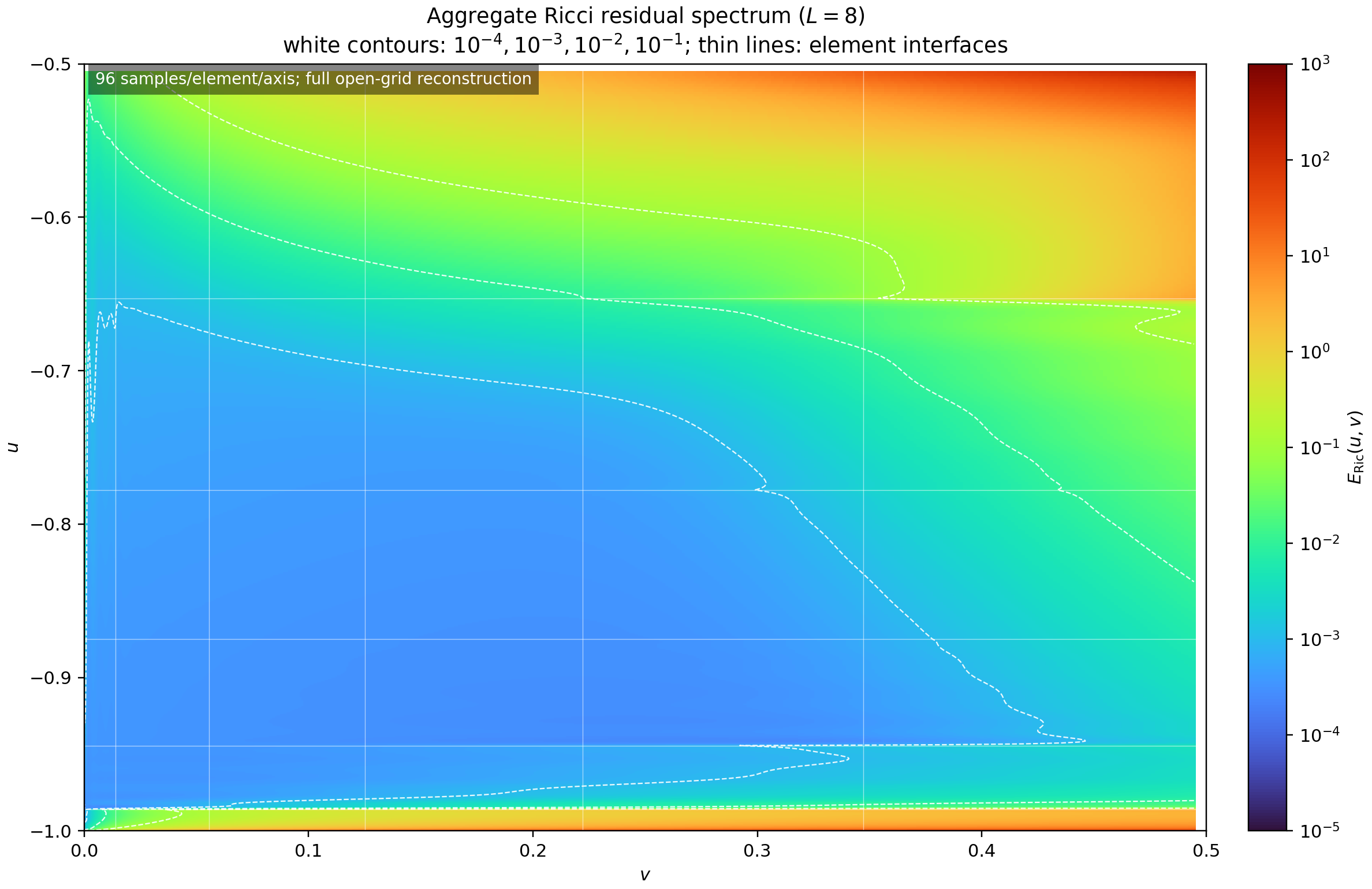}
  \caption{Spectrum of the aggregate Ricci residual
  $E_{\mathrm{Ric}}(u,v)$.  Colors show the elementwise spectral
  interpolation of $\log_{10}E_{\mathrm{Ric}}$ on the open coordinate
  grid; white contours mark $10^{-4}$, $10^{-3}$, $10^{-2}$, and $10^{-1}$,
  and the thin lines are coordinate-element interfaces.}
  \label{fig:results-exp5-residual-spectrum}
\end{figure}

Figure \ref{fig:results-exp5-residual-spectrum} includes the unprotected
layers of the open grid.  The extreme null-curvature components can diverge
at the initial corner, but they are not differentiated to obtain this
residual.  The large unprotected values instead expose errors in the
first-order differentiation near the fractional-power endpoints and element
interfaces.  They are not used as an interior accuracy statistic; the
protected values are reported in Table~\ref{tab:results-exp5-residual}.

% \clearpage
\section{Results for Einstein Scalar-field Equations}

We test the Einstein--scalar-field iteration in three complementary regimes.
Experiment 6 uses the explicit Fisher--Janis--Newman--Winicour family to
measure the section-metric error directly.  Experiment 7 replaces spherical
symmetry by smooth angular lapse, shift, shear, and scalar data; without an
explicit reference spacetime, its accuracy is measured by the independently
differentiated first-order curvature residual
$\Rc-\dd\phi\otimes\dd\phi$.  Experiment 8 strengthens the scalar pulse,
covers a curved characteristic region by an overlapping inner atlas, locates
a trapped region, and reconstructs the associated apparent horizon as a tube
of marginally outer trapped surfaces.

\subsection{Fisher--Janis--Newman--Winicour benchmarks}

The Fisher--Janis--Newman--Winicour family is the standard static,
spherically symmetric solution of the Einstein--massless-scalar equations.
For $\sigma>0$, $0<\nu\leq1$, and
\begin{equation}
  f(r)=1-\frac{\sigma}{r},
  \qquad c_\nu=\sqrt{\frac{1-\nu^2}{2}},
\end{equation}
the metric and scalar field in static coordinates are
\begin{align}\label{eq:results-jnw-static}
  \bg_{\mathrm{JNW}}
  &=-f^\nu\dd t^2+f^{-\nu}\dd r^2
    +r^2f^{1-\nu}\roundg,\\
  \phi&=c_\nu\log f.
\end{align}
They satisfy $\Rc_{\mu\nu}=\partial_\mu\phi\,\partial_\nu\phi$ with the
normalization used here.  Introduce an optical radius and null coordinates by
\begin{equation}
  \frac{\dd r_*}{\dd r}=f^{-\nu},\qquad
  t=u+v,\qquad r_*=v-u-\frac12,
  \qquad r(0)=2\sigma.
\end{equation}
Then \eqref{eq:results-jnw-static} takes the double-null form
\begin{equation}\label{eq:results-jnw-double-null}
  \bg_{\mathrm{JNW}}
  =-4\Omega^2\dd u\dd v+\slg,\qquad
  \Omega^2=f^\nu,\qquad
  \slg=r^2f^{1-\nu}\roundg,\qquad b=0.
\end{equation}

We set $\sigma=1$ and use
\begin{equation}
  \nu=0.99,\ 0.8,\ 0.5,\ 0.2
\end{equation}
on the numerical rectangle
\begin{equation}\label{eq:results-exp6-domain}
  -1\leq u\leq-\frac12,\qquad 0\leq v\leq\frac12.
\end{equation}
Thus $0\leq r_*\leq1$ and $2\leq r\leq2.89832$ across the four cases, so the domain remains
strictly outside the curvature singularity $r=\sigma$.  The characteristic
traces are extracted from \eqref{eq:results-jnw-double-null}, while the
interior is generated by eight Picard sweeps.  Coordinate refinements use
two, four, and eight degree-six elements in each null direction; the finest
grid has $49\times49$ coordinate nodes.  A separate $\nu=1$ control reduces
to Schwarzschild with mass $M=\sigma/2=1/2$ and has identically vanishing
scalar variables.

As in the vacuum benchmarks, define the pointwise section-metric error by
\begin{equation}\label{eq:results-exp6-metric-error}
  E_g(q)=\norm{\slg_{\mathrm{num}}(q)
    -\slg_{\mathrm{exact}}(q)}_{\mathrm F}.
\end{equation}
Table \ref{tab:results-exp6-metric-error} gives its statistics over all sphere
and coordinate nodes of the finest grid.

\begin{table}[htbp]
  \centering
  \caption{Final section-metric error for the Fisher--JNW benchmarks on the
  $49\times49$ coordinate grid.}
  \label{tab:results-exp6-metric-error}
  \begin{tabular}{crrr}
    \toprule
    $\nu$ & $\operatorname{median}E_g$ & $\operatorname{mean}E_g$
      & $\max E_g$ \\
    \midrule
    $0.99$ & $7.211\times10^{-13}$ & $7.372\times10^{-13}$ & $1.717\times10^{-12}$ \\
$0.80$ & $1.856\times10^{-12}$ & $1.740\times10^{-12}$ & $3.751\times10^{-12}$ \\
$0.50$ & $7.676\times10^{-12}$ & $7.475\times10^{-12}$ & $2.790\times10^{-11}$ \\
$0.20$ & $3.576\times10^{-11}$ & $4.513\times10^{-11}$ & $5.641\times10^{-10}$ \\
    \bottomrule
  \end{tabular}
\end{table}

The error increases as $\nu$ moves away from the Schwarzschild limit and the
scalar field becomes stronger, but even the $\nu=0.2$ case retains a maximum
absolute metric error below $5.7\times10^{-10}$.

\subsection{Nonspherical scalar characteristic data}

The seventh experiment prescribes nonspherical characteristic data on
\begin{equation}\label{eq:results-exp7-domain}
  -1\leq u\leq-\frac12,
  \qquad 0\leq v\leq0.04.
\end{equation}
Let $r=-u$ on the incoming initial hypersurface, and let $(x,y,z)$ be
Cartesian coordinates restricted to the unit sphere.  Define the smooth fields
\begin{equation}
  Y=\frac{3z^2-1}{2},\qquad V=\nabla_{\roundg}(xz),\qquad
  Q=\bigl(\nabla^2_{\roundg}(x^2-y^2)\bigr)^{\mathrm{tf}},
  \qquad P_0=\frac{8\sqrt2}{5}.
\end{equation}
The free incoming data are
\begin{equation}\label{eq:results-exp7-incoming-data}
  \slg(u,0)=r^2\roundg,\qquad
  \Omega^2(u,0)=r^{1/4}(1+0.01Y),\qquad b(u,0)=0.05V.
\end{equation}
On the outgoing initial hypersurface we prescribe
\begin{align}\label{eq:results-exp7-outgoing-data}
  \Omega^2(-1,v)&=1+0.01Y,\\
  (\Omega e_4\phi)(-1,v)&=P_0+v^{0.1},\\
  \Omega\chih(-1,v)&=0.1v^{0.1}\mathcal T_Q[\slg(-1,v)].
\end{align}
The normalizations are fixed analytically, independently of the sphere grid.
The free metric, log-lapse, shift, and tensor profiles are represented in the
prescribed angular Galerkin spaces before completing the face constraints.
At the corner $\phi=0$, $\Omega\tr\chi=1.6$, $\zeta=0$, and
$\Omega\omega=0$.  The positive branch of $\Omega e_3\phi$ and all remaining
face fields are determined by the Einstein--scalar characteristic
constraints.  The functions in \eqref{eq:results-exp7-incoming-data}--
\eqref{eq:results-exp7-outgoing-data} are globally smooth in angle, while the
power coordinate
\begin{equation}
  \tau=-\log(-u),\qquad s=(v/0.04)^{0.1}
\end{equation}
resolves their prescribed $v^{0.1}$ corner behavior.

There is no explicit spacetime for comparison.  After independent angular resampling and coordinate overgrid
refinement we therefore use
\begin{equation}\label{eq:results-ese-curvature-error}
  E_{\mathrm{Ric}}(u,v)
  =\norm{\Rc-\dd\phi\otimes\dd\phi}_{L^2(S_{u,v})}.
\end{equation}
Here the Ricci components are computed from first null derivatives of the
resampled weighted connection variables, with metric--connection and scalar
consistency and the scalar wave residual checked separately.  The positive norm squares the null
components with weights $1,1,2,1,1$ for $33,44,34,3A,4A$, respectively,
and uses the full tensor norm for $AB$; it is not a Lorentzian contraction.
For coordinate comparison, the audit uses the union of the two- and
three-element source partitions, with degrees 11 and 14 in $\tau$ and $s$.
All three solutions are therefore evaluated on the same $45\times57$ grid.
The protected region removes three nodes from each outer endpoint and each
audit-element interface, requires $s\geq0.6$, and contains 198 sections.
The scalar, vector, and tensor fields are transferred in their typed spaces;
for the central angular discretization the transfer degree is 16 on 362
independent sphere points. Table
\ref{tab:results-exp7-curvature-error} shows coordinate convergence with the
central $L=8$, $W=16$, 350-point angular discretization held fixed.

\begin{table}[htbp]
  \centering
  \caption{Protected independent curvature residual for Experiment 7.  Grid
  sizes refer to the evolved state before independent overgrid resampling.}
  \label{tab:results-exp7-curvature-error}
  \begin{tabular}{crrr}
    \toprule
    Coordinate grid & $\operatorname{median}E_{\mathrm{Ric}}$
      & $\operatorname{mean}E_{\mathrm{Ric}}$
      & $\max E_{\mathrm{Ric}}$ \\
    \midrule
    $13\times19$ & $5.947\times10^{-3}$ & $9.907\times10^{-3}$ & $6.251\times10^{-2}$ \\
$17\times23$ & $2.677\times10^{-3}$ & $3.706\times10^{-3}$ & $1.361\times10^{-2}$ \\
$25\times34$ & $1.550\times10^{-3}$ & $2.676\times10^{-3}$ & $1.070\times10^{-2}$ \\
    \bottomrule
  \end{tabular}
\end{table}

\begin{figure}[htbp]
  \centering
  \includegraphics[width=0.8\textwidth]{
    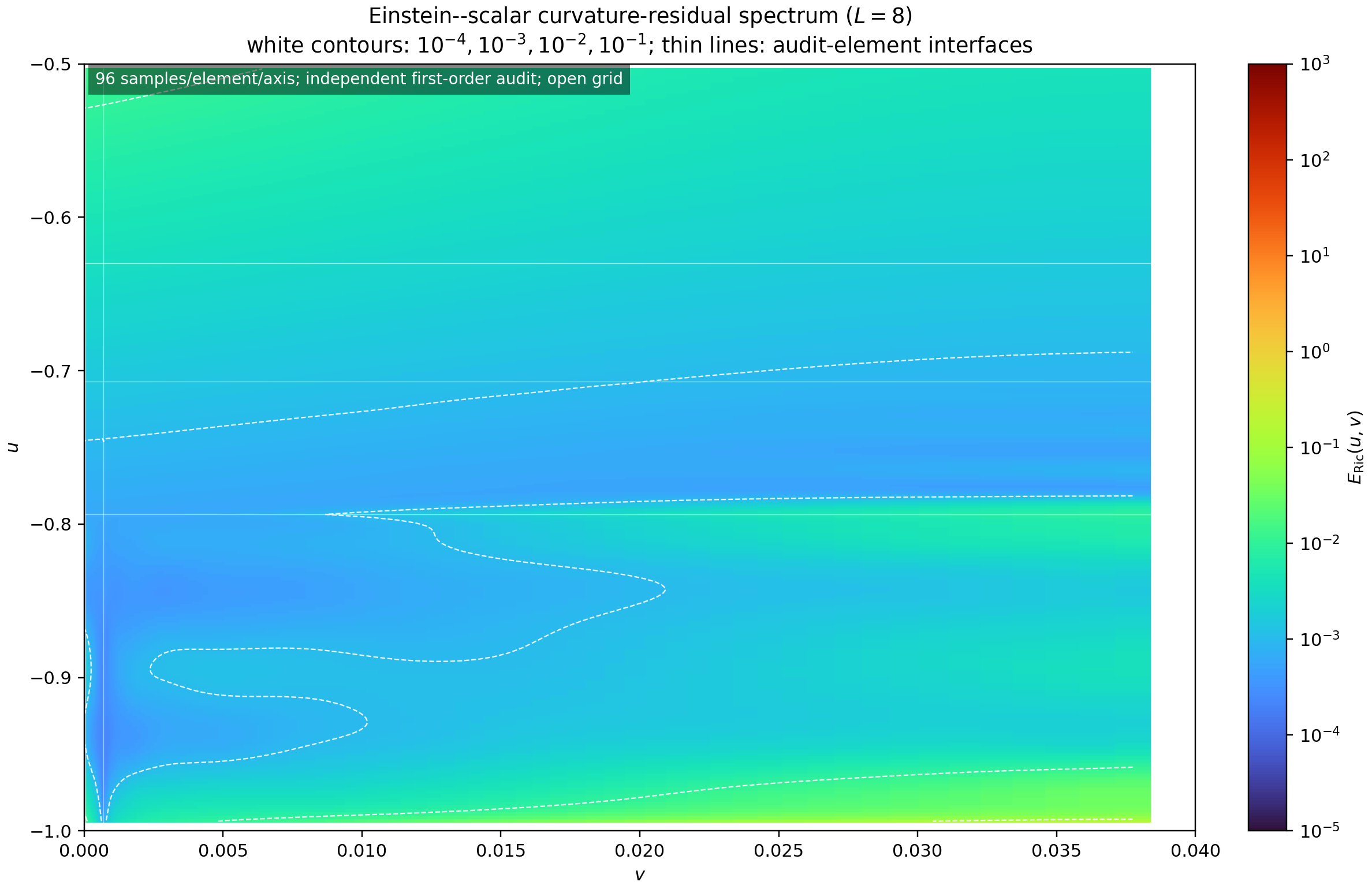}
  \caption{Spectrum of the independent Einstein--scalar curvature residual
  $E_{\mathrm{Ric}}(u,v)$ for the finest coordinate run.  Colors show
  elementwise spectral interpolation of $\log_{10}E_{\mathrm{Ric}}$; white
  contours mark $10^{-4}$, $10^{-3}$, $10^{-2}$, and $10^{-1}$, and the thin
  lines mark audit-element interfaces.}
  \label{fig:results-exp7-curvature-spectrum}
\end{figure}

All three coordinate runs complete six Picard sweeps, and the protected
maximum decreases from $6.25\times10^{-2}$ to $1.07\times10^{-2}$.
At fixed central coordinate resolution, the sum of the protected Einstein
and scalar-wave residual maxima decreases from $3.99\times10^{-2}$ through
$1.55\times10^{-2}$ to $9.79\times10^{-3}$ under angular refinement.
Figure
\ref{fig:results-exp7-curvature-spectrum} displays the finest-coordinate
audit.  The plot includes the full open overgrid to expose the unresolved
$v^{0.1}$ endpoint layer, whereas the statistics in Table
\ref{tab:results-exp7-curvature-error} use only the protected region.

\subsection{Scalar-pulse formation of a trapped region and apparent horizon}

Experiment 8 keeps the incoming spherical geometry
\begin{equation}\label{eq:results-exp8-incoming-data}
  \slg(u,0)=(-u)^2\roundg,\qquad
  \Omega^2(u,0)=(-u)^{1/4},\qquad b(u,0)=0,
\end{equation}
but selects the negative branch of the incoming scalar constraint.  At the
corner,
\begin{equation}
  \phi=0,\qquad \Omega\tr\chi=1.6,\qquad
  \Omega\tr\chib=-2,\qquad
  (\Omega e_4\phi)(-1,0)=-2.2627416998.
\end{equation}
Use the axisymmetric tensor
$Q_{\rm ax}=-(\nabla^2_{\roundg}z^2)^{\mathrm{tf}}$ and set
\begin{equation}
  F(v)=\left(\frac{v}{v+0.01}\right)^{0.1},
\end{equation}
the outgoing scalar and shear perturbations are
\begin{align}\label{eq:results-exp8-outgoing-data}
  e_4\phi(-1,v)-e_4\phi(-1,0)&=-2F(v),\\
  \Omega\chih(-1,v)&=0.1F(v)
     \mathcal T_{Q_{\rm ax}}[\slg(-1,v)].
\end{align}
Here $\Omega(-1,v)=1$, so the first line is also the corresponding statement
for the stored variable $\Omega e_4\phi$.  The characteristic constraints
determine the remaining data.  On the initial faces the outgoing expansion
is still positive, with minimum $0.880366$, while $\tr\chib$ is negative.
Thus the initial data themselves contain no trapped section.

The evolution region is changed to
\begin{equation}\label{eq:results-exp8-curved-domain}
  \mathcal D=\left\{(u,v):-1\leq u\leq-0.05,\quad
  0\leq v\leq\min\left(0.1(-u)^{25/24},0.05\right)\right\}.
\end{equation}
This cap follows the focusing scale and avoids evolving the excluded part of
a bounding rectangle through the double-null caustic.  The numerical solver
remains rectangular: seven nested characteristic rectangles, each wholly
contained in $\mathcal D$, form an inner staircase atlas covering $94.6694\%$
of its coordinate area.  Adjacent patches are compared in their overlaps;
the largest absolute discrepancy among the checked primitive fields is
$5.74\times10^{-7}$, while the section-metric discrepancy is below
$1.53\times10^{-8}$.

Table \ref{tab:results-exp8-curvature-error} applies
\eqref{eq:results-ese-curvature-error} on the same $s\geq0.6$ protected region
used above.  These rows compare different atlas patches; the separate angular
control tests the coordinate-refined anchor near $v=0.04$.

\begin{table}[htbp]
  \centering
  \caption{Protected independent curvature residual on the seven production
  patches of the Experiment 8 inner atlas.}
  \label{tab:results-exp8-curvature-error}
  \small
  \begin{tabular}{lrrrr}
    \toprule
    $u_{\rm R}$ & $v_{\rm cap}$
      & $\operatorname{median}E_{\mathrm{Ric}}$
      & $\operatorname{mean}E_{\mathrm{Ric}}$
      & $\max E_{\mathrm{Ric}}$ \\
    \midrule
    $-0.52$ & $0.050000$ & $2.365\times10^{-4}$ & $2.097\times10^{-3}$ & $3.439\times10^{-2}$ \\
$-0.45$ & $0.043527$ & $2.141\times10^{-4}$ & $1.605\times10^{-3}$ & $1.915\times10^{-2}$ \\
$-0.35$ & $0.033502$ & $1.418\times10^{-4}$ & $1.096\times10^{-3}$ & $1.268\times10^{-2}$ \\
$-0.25$ & $0.023597$ & $8.421\times10^{-5}$ & $7.923\times10^{-4}$ & $6.761\times10^{-3}$ \\
$-0.15$ & $0.013860$ & $5.655\times10^{-5}$ & $6.318\times10^{-4}$ & $4.317\times10^{-3}$ \\
$-0.10$ & $0.009085$ & $3.761\times10^{-5}$ & $5.097\times10^{-4}$ & $4.923\times10^{-3}$ \\
$-0.05$ & $0.004413$ & $9.847\times10^{-6}$ & $1.451\times10^{-4}$ & $2.446\times10^{-3}$ \\
    \bottomrule
  \end{tabular}
\end{table}

Here $u_{\rm R}$ and $v_{\rm cap}$ are the right and upper boundaries of a
patch.  The protected mask requires $s\geq0.6$ and removes three overgrid
nodes at outer boundaries and at every source-element interface.  The table
therefore measures the independently reconstructed Einstein residual away
from both the fractional-power corner layer and spectral differentiation
interfaces.

Finally, we resample both null expansions independently on 1000 sphere points
and search for a coordinate section satisfying
\begin{equation}
  \sup_{S_{u,v}}\tr\chi<0,
  \qquad
  \sup_{S_{u,v}}\tr\chib<0.
\end{equation}
The atlas contains such sections across several patches, producing a sampled
trapped region rather than a single endpoint candidate.  Direct four-metric
evaluation at the reference section
\begin{equation}\label{eq:results-exp8-trapped-section}
  (u,v)=(-0.4696271025,0.04),
\end{equation}
gives
\begin{equation}
  \sup_{S_{u,v}}\theta_+=-0.0802152,
  \qquad
  \sup_{S_{u,v}}\theta_-=-4.28011.
\end{equation}
Both null expansions are therefore strictly negative on this entire
two-sphere.  The independent $L=7$, $W=14$, 300-point angular control gives
$-0.0802016$ and $-4.27999$, respectively.

To reconstruct the boundary of the trapped region, on each incoming cone
$v=v_0$ we solve the angular graph equation
\begin{equation}\label{eq:results-exp8-mots-equation}
  \theta_+[u=h(\theta^A),v_0]=0.
\end{equation}
The expansion is calculated from the four-metric connection and graph
embedding and does not reuse the evolved expansion variable.  A section is
retained only if the nonlinear solve succeeds within the patch interior,
$\norm{\theta_+}_{L^\infty}\leq10^{-5}$, $\theta_-$ is strictly negative,
and small inward and outward constant displacements bracket $\theta_+$.
The search starts from the outermost detected constant-section bracket;
this does not establish outermostness among arbitrary angular graphs.

\begin{figure}[htbp]
  \centering
  \includegraphics[width=0.9\textwidth]
    {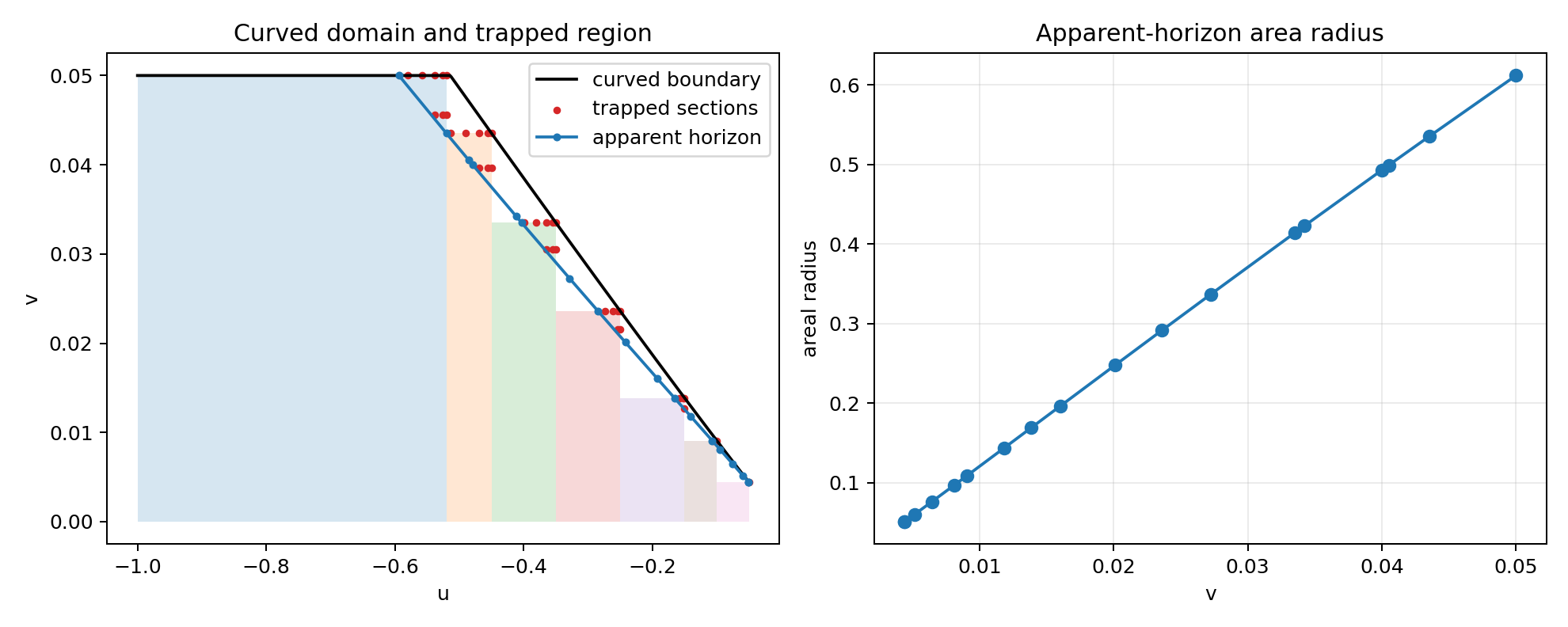}
  \caption{Experiment 8 curved characteristic region, inner rectangular
  atlas, sampled trapped sections, and reconstructed apparent horizon. Shading
  shows the atlas union, red points are trapped sections, and the blue curve
  joins the reconstructed MOTSs. The right panel shows their areal radii.}
  \label{fig:results-exp8-curved-domain-horizon}
\end{figure}

The resulting apparent horizon contains 18 verified MOTSs spanning
$0.00441\leq v\leq0.05$.  The mean graph location runs from
$u=-0.0504104$ to $u=-0.5938472$, while the areal radius increases from
$0.0512842$ to $0.6118550$.  Over the complete tube,
\begin{equation}
  \max_v\norm{\theta_+}_{L^\infty(S)}=8.47\times10^{-6},
  \qquad \max_v\sup_{S_{h(v),v}}\theta_-=-3.38460,
\end{equation}
and the maximum peak-to-peak angular deformation of $h$ is
$1.33\times10^{-4}$.  At $v=0.04$, the degree-five coordinate-refined MOTS
has
\begin{equation}
  \overline h=-0.479401999,\qquad
  |S|=3.05516039,\qquad r_S=0.493073969.
\end{equation}
Refining the characteristic grid changes $h$ by
$1.50\times10^{-7}$ in maximum norm and changes the area by
$6.54\times10^{-7}$ relatively.  The 300-point angular control changes the
mean location by $7.80\times10^{-10}$ and the area by
$3.29\times10^{-9}$ relatively.  A degree-three horizon trace agrees in mean
location to $2.97\times10^{-8}$ but leaves
$\norm{\theta_+}_{L^\infty}$ as large as $9.40\times10^{-3}$; degree four is
therefore required to resolve the nonspherical correction.

% \clearpage
\section[Conclusion and Outlook]{\hspace{0.4em}Conclusion and Outlook}%
\label{sec:conclusion}

We have developed a first-order Picard iteration for the characteristic
initial value problem for the Einstein vacuum and Einstein--massless-scalar
equations in double-null gauge.  The construction follows the geometric
dependency structure of the equations and combines characteristic constraint
solves with spectral elements in the two null coordinates, pole-free angular
operators, and independent first-order residual audits with separate
metric--connection consistency checks.  This provides a unified
numerical framework for exact benchmarks and for characteristic data with no
explicit interior solution.

The regular exact tests cover vacuum Schwarzschild and Kerr geometries and the
scalar Fisher--JNW family.  Metric errors remain small in regular domains and
across a Schwarzschild horizon.  The Schwarzschild-interior experiment also
identifies the present method's loss of accuracy as the curvature singularity
is approached.  Beyond the explicit families, the vacuum calculations evolve
a strong nonspherical outgoing perturbation and crossed characteristic data
with singular limiting initial curvature.  The Einstein--scalar calculations
further demonstrate
coordinate convergence for nonspherical, low-regularity characteristic data
on the protected numerical region.

The final experiment gives numerical evidence for a trapped region and
reconstructs its apparent horizon.  On the curved characteristic domain, both
angular discretizations identify the section
\begin{equation}
  (u,v)=(-0.4696271025,0.04),
\end{equation}
where the two future null expansions are strictly negative on the entire
sphere.  Eighteen independently reconstructed MOTSs form a sampled
apparent-horizon tube over $0.00441\leq v\leq0.05$.  Coordinate refinement
changes its $v=0.04$ graph by $1.50\times10^{-7}$ in maximum norm, while an
independent higher-angular-band evolution changes its mean location by
$7.80\times10^{-10}$.  The distinction between exact metric errors,
independently evaluated curvature residuals, trapped-section sign tests, and
the graph MOTS equation remains essential when interpreting the experiments.

Three directions are particularly important for future work.  First, the
angular MOTS solve should be extended with adaptive coordinate refinement
toward the singular endpoint and across alternative foliations, thereby
resolving the anisotropic apparent horizon more completely.  Second, varying
the scalar-pulse and anisotropic-perturbation
amplitudes would reveal the threshold for trapping and the dependence of the
first trapped location on the initial data.  Third, improved coordinates,
adaptive refinement, and more robust iteration strategies are needed to
improve the approximation near curvature singularities.  These developments
would extend the present double-null framework from a local numerical
construction toward a more precise study of anisotropic black-hole formation
and singular spacetime geometry.

\section*{Acknowledgments}
S.W. is supported by the NUS President Graduate Fellowship. 

\section*{Code availability}
The source code, experiment configurations, and instructions for reproducing
the numerical experiments are available at\\
\url{https://github.com/Shengrong-Wu/Numerical-Einstein-Equations}.

\bibliographystyle{plain}
\bibliography{ref}

\end{document}